\documentclass[12pt]{article}
\pdfoutput=1
\usepackage{amssymb}
\usepackage{amsmath}
\usepackage{bm} 
\usepackage{graphicx}
\usepackage{color}
\usepackage{xcolor}
\usepackage{dsfont}
\usepackage{hyperref}
\definecolor{nicered}{rgb}{0.7,0.1,0.1}
\definecolor{nicegreen}{rgb}{0.1,0.5,0.1}
\hypersetup{colorlinks,citecolor= nicegreen,linkcolor= nicered}
\begin{document}
\begin{titlepage}

  \newcommand{\AddrUdeA}{{\sl \small Instituto de Fisica, 
      Universidad de Antioquia,\\A.A. 1226, Medellin, Colombia}}
  \newcommand{\AddrLiege}{{\sl \small IFPA, Dep. AGO, 
      Universite de Liege, Bat B5,\\ \small \sl Sart
      Tilman B-4000 Liege 1, Belgium}}
  \newcommand{\AddrSISSA}{{\sl \small SISSA and INFN, Sezione
      di Trieste,\\ Via Bonomea 265, 34136 Trieste, Italy}}
\vspace*{0.5cm}
\begin{center}
  \textbf{\large LSP sneutrino novel decays
  }\\[15mm]
  D. Aristizabal Sierra$^{a,}$\footnote{e-mail address:
    daristizabal@ulg.ac.be}, D. Restrepo$^{b,}$\footnote{e-mail address:
    restrepo@udea.edu.co}, S. Spinner$^{c,}$\footnote{e-mail address:
    sspinner@sissa.it} 
  \vspace{1cm}\\
  $^a$\AddrLiege.\vspace{0.4cm}\\
  $^b$\AddrUdeA.\vspace{0.4cm}\\
  $^c$\AddrSISSA.
\end{center}
\vspace*{0.2cm}

\begin{abstract}
  In bilinear R-parity violation (BRpV), in which the superpotential
  includes a bilinear term between the lepton doublet and the up-type
  Higgs superfields, a sneutrino LSP can decay into pairs of heavy
  standard model states ($W$s, $Z$s, tops or Higgs bosons),
  neutrinos or different-flavor charged lepton modes ($l_il_j$). These
  finals states can dominate over the traditionally considered bottom
  pair final state and would lead to unique and novel
  supersymmetric signals: multileptons events or pairs of heavy standard model fields. We
  investigate this possibility and find that the branching ratio into
  these states dominates when the bilinear term is much smaller than
  the sneutrino vacuum expectation value for a given sneutrino
  flavor. When BRpV is the only source of neutrino masses these decays
  can only dominate for one of the sneutrino generations. Relaxing
  this constraint opens these channels for all three generations.
\end{abstract}
\end{titlepage}
\setcounter{footnote}{0}
\section{Introduction}
\label{sec:intro}

As the large hadron collider (LHC) continues to successfully probe the
nature of electroweak symmetry breaking (a recent breakthrough being
the discovery of a Higgs candidate~\cite{ATLAS, CMS}), a solution to
the gauge hierarchy problem, if it exists, has evaded our efforts so
far. An elegant candidate for such a solution and one which addresses
several other open issues as well, \textit{e.g.} dark matter and gauge
coupling unification, is supersymmetry (SUSY). Because of its
theoretical appeal, it is important to understand all the guises that
SUSY may adopt in order to recognize it if it is produced at the LHC.

An important open issue in SUSY, which has strong ramifications for
its LHC phenomenology, as well as its cosmology, is the gauge
invariance of lepton and baryon number violating interactions. Aside
from introducing many new unknown parameters, these interactions also
lead to rapid proton decay. The most common solution to address this
issue is the imposing of a discrete symmetry, R-parity, defined as
$R_p \equiv (-1)^{3(B-L) +2S}$ (see \cite{Barbier:2004ez} for a
review). This forbids all tree-level lepton and baryon number
violating terms and also causes the lightest supersymmetric particle
(LSP) to be stable and therefore neutral. This LSP can then play the
role of dark matter and would manifest itself at colliders as
missing energy. However, proton decay requires the elimination of only
the lepton number or baryon number violating terms and from a
theoretical perspective, R-parity can be \textit{ad hoc}. Furthermore,
an open-mindedness to possible signals at the LHC should push us to
consider alternatives. Finally, when R-parity is violated the
stringent constraints on superpartner masses, derived from negative
collider searches for missing energy events at the Tevatron and the LHC
\cite{Chatrchyan:2012te,Aad:2011cwa}, can be relaxed\footnote{See
  ref. \cite{Carpenter:2007zz,Carpenter:2008sy,Graham:2012th} for
  details.}.

A systematic study of all possible R-parity violating terms and their
effects on phenomenology is an arduous task and furthermore, one would
like a mechanism for understanding why proton decay is significantly
suppressed. A well-motivated solution to both of these issues are
models which can predict the fate of R-parity. A natural framework for
this endeavor is in the context of $U(1)_{B-L}$ symmetries,
see~\cite{AM, Hayashi, Mohapatra} for early
examples\footnote{Horizontal symmetries $U(1)_X$ can be also used to
  construct models where the RpV couplings, arising from effective
  operators, are intrinsically small
  \cite{Mira:2000gg,Sierra:2009zq}.}. While some $B-L$ models predict
R-parity conservation~\cite{Goran1,Goran2}, the most minimal ones (in
terms of particle content) require R-parity violation (RpV)~\cite{AM,
  Mohapatra, Rp1,Rp2, Ambroso:2009jd}. Even some non-minimal models
prefer RpV from considerations of the renormalization group evolution
of the soft masses~\cite{Hayashi, Fate}. These and other models of
spontaneous R-parity violation, such as~\cite{Masiero:1990uj,
  Buchmuller:2007ui}, have the common feature that they can be
described, in an effective field theory way, by bilinear R-parity
violation (BRpV): the only R-parity violating terms are the mixings
between the lepton doublets and up-type Higgs doublet in the
superpotential. This makes BRpV a powerful tool for studying possible
signatures of spontaneous RpV. Furthermore, proton decay is highly
suppressed and it is important to mention that a gravitino LSP can be
a dark matter candidate in such models~\cite{Buchmuller:2007ui,
  Takayama:2000uz}.

Once R-parity is broken the LSP is no longer stable and therefore
astrophysical constraints on its nature do not apply
\cite{Dimopoulos:1989hk}. Accordingly, from a purely phenomenological
point of view any superpartner can be the LSP, and studies of the
different possibilities in BRpV models (and of sleptons and sneutrinos
in general models~\cite{Bartl:2003uq,Aristizabal Sierra:2004cy}) and
their relation to neutrino masses and mixings have been carried out in
e.g. \cite{Porod:2000hv,Hirsch:2003fe,Roy:1996bua,Mukhopadhyaya:1998xj}.

In this paper we extend upon previous results by considering sneutrino
LSP\footnote{A sneutrino NLSP with a gravitino LSP would not change
  our phenomenological results.}  decays in BRpV into
``non-conventional'' modes: pairs of heavy standard model (SM) states ($W^+W^-$,
$Z^0Z^0$, $h^0h^0$, $t\bar t$),
invisible modes ($\nu\nu$) and different-flavor charged lepton final states ($l_i^+l_j^-$, $i\neq
j$). Such states, to our knowledge, had not received much
attention before despite the fact that they can dominate sneutrino
decays and can yield unique and unanticipated SUSY
signals\footnote{Gauge boson and top quark pair production through a
  sneutrino resonance in trilinear RpV was studied in
  \cite{BarShalom:2001ew} while the full set of heavy SM modes have
  been mentioned in BRpV models in
  \cite{Grossman:1998py}}. Specifically, neutrino masses force RpV to
be small so its only effect is on the decay of the LSP. Therefore, for
a sneutrino LSP, every SUSY event will eventually decay into two
sneutrinos which could then decay into one of the aforementioned
states. As we will show, this typically translates into into different-flavor charged leptons
or heavy SM modes, depending on the sneutrino generation.

The main goal of this paper is to study the sneutrino decays into
these ``non-conventional'' modes and to show that they can dominate
over the traditionally considered $b \bar b$ final state. We find that
the latter dominate roughly when the BRpV term, $\epsilon_i$ is
smaller than the vacuum expectation value (vev) of the sneutrino,
$v_i$, for a given flavor of sneutrino, $i$. In the case when BRpV is
the only source of neutrino masses, this possibility can only hold for
one sneutrino flavor, however, when this assumption is relaxed, it can
hold true for all three generations. Therefore, if two or more
generations of sneutrinos decay via RpV, it might be possible to rule
out BRpV as the sole generator of neutrino masses.

The rest of the paper is organized as follows. In section
\ref{sec:generalities} we discuss the generalities of the BRpV model,
in particular those related with the neutral scalar sector. In section
\ref{sec:sneutrino-mix-analyR} we derive formulas for BRpV induced
mixings. In section~\ref{sec:sneutrino-decays-analyticalR} we write
the relevant couplings for BRpV sneutrino decays, give analytical
formulas for the different partial decay widths, analyze the
constraints on parameter space enforced by neutrino data and present
our results. In section~\ref{sec:conclusions} we summarize and present
our conclusions.

\section{Bilinear R-parity violation}
\label{sec:generalities}
In what follows we will briefly describe the main features of the
bilinear R-parity breaking model, in particular those related with the
neutral scalar sector. We shall closely follow the notation used in
\cite{Hirsch:2000ef} and assume that all parameters are
real\footnote{This simplification does not affect our main
  conclusions.}. Throughout the text matrices will be denoted in
bold-face.

In addition to the MSSM R-parity conserving superpotential (where we have suppressed relevant indices):
\begin{equation}
	W_\text{MSSM} = \boldsymbol{h^U} \hat Q \hat H_u \hat u^c + \boldsymbol{h^D} \hat Q \hat H_d \hat d^c
	+ \boldsymbol{h^E} \hat L \hat H_d \hat e^c + \mu \hat H_u \hat H_d
\end{equation}
the bilinear R-parity breaking model also contains the following terms:
\begin{equation}
  \label{eq:super-potential}
  W_\text{BRpV}=\epsilon_{\alpha\beta}\,\epsilon_i\,
  \hat L_i^\alpha\,\hat H_u^\beta\,,
\end{equation}
where $\epsilon_{\alpha\beta}$ is the $SU(2)$ completely antisymmetric
tensor, $i=1,2,3$ runs over the SM fermion generations and
$\epsilon_i$ is the R-parity and lepton number breaking bilinear parameter with
units of mass. Consistency then requires a new set of soft
SUSY breaking terms in the scalar potential, namely
\begin{equation}
  \label{eq:V-BRpV-soft}
  V_\text{BRpV}= B_i\,\epsilon_i \, \epsilon_{\alpha\beta}\,\tilde L_i^\alpha\,H_u^\beta\,.
\end{equation}
Neglecting soft flavor mixing, the scalar potential relevant for neutral scalars is
\begin{align}
  \label{eq:full-SSB-scalar-potential}
  V&\supset
  (m_{H_d}^2+\mu^2)H_d^\dagger H_d 
  + (m_{H_u}^2+\mu^2)H_u^\dagger H_u
  + m_{L_i}^2\tilde L_i^\dagger \tilde L_i
  \nonumber\\
  &
  +\frac{1}{8}g_Z^2(H_u^\dagger H_u - H_d^\dagger H_d - \tilde L_i^\dagger \tilde L_i)^2
  + |\boldsymbol{\epsilon}|^2H_u^\dagger H_u
  + \epsilon_i\,\epsilon_j\,\tilde L_i^\dagger \tilde L_j
  \nonumber\\
  &
  + \left(
	  - \mu\,\epsilon_i\tilde L_i^\dagger H_d
	  - B\mu\epsilon_{\alpha\beta}H_d^\alpha H_u^\beta
	  + B_i\,\epsilon_i \, \epsilon_{\alpha\beta}\,\tilde L_i^\alpha\,H_u^\beta + \text{H.c.}
  \right),
\end{align}
with $g_Z^2=g^2+g^{\prime 2}$ and
$\boldsymbol{\epsilon}^T=(\epsilon_1,\epsilon_2,\epsilon_3)$. Electroweak
symmetry is broken once the Higgs and slepton acquire
a vev, $\langle H_{d,u}\rangle=v_{d,u}/\sqrt{2} $ and $\langle \tilde
L_i\rangle=v_i/\sqrt{2} $, with $v=(v_u^2+v_d^2+\sum_{i=1,2,3}v^2_i)^{1/2}\simeq
246$~GeV and $M_Z^2=g_Z^2v^2/4$. The doublets are parameterized as
\begin{equation}
  \label{eq:higgs-slepton-doublets}
  H_d=
  \begin{pmatrix}
    H_d^0\\
    H_d^-
  \end{pmatrix}\,,\quad
  H_u=
  \begin{pmatrix}
    H_u^+\\
    H_u^0
  \end{pmatrix}\,,\quad
    \tilde L_i=
  \begin{pmatrix}
    \tilde L_i\\
    l^-_i
  \end{pmatrix}\,,
\end{equation}
with the neutral components given by
\begin{equation}
  \label{eq:neutral-fields}
  H^0_d=\frac{1}{\sqrt{2}}(\sigma_d^0+i\varphi_d^0+v_d)\,,\quad
  H^0_u=\frac{1}{\sqrt{2}}(\sigma_u^0+i\varphi_u^0+v_u)\,,\quad
  \tilde L_i=\frac{1}{\sqrt{2}}(\tilde\nu^R_i+i\tilde\nu^I_i+v_i)\,.
\end{equation}
Here we introduce the notation $\tilde\nu^{R,I}$ to differentiate the
CP-even sneutrinos from the CP-odd.

In the basis $(S^0)^T=(\sigma_d^0,\sigma_u^0,\tilde\nu_i^R)^T$ the
linear part of the neutral scalar potential can be written as
\begin{equation}
  \label{eq:vlinear}
  V_\text{linear}(S^0_i)=\sum_{a=1,\cdots,5} t_a\,S_a^0\,,
\end{equation}
where the $t_a$'s are the so-called tadpoles. At tree-level these
are (see eqs. (\ref{eq:full-SSB-scalar-potential}) and
(\ref{eq:vlinear}))
\begin{align}
  \label{eq:tree-lvel-tadpoles}
  t_d^{(0)}&=(m_{H_d}^2+\mu^2)v_d + D v_d - 
  \mu(B v_u + \boldsymbol{\epsilon}\cdot\boldsymbol{v})\,,\nonumber\\
  t_u^{(0)}&=-B\mu v_d + (m_{H_u}^2+\mu^2)v_u - D v_u
  +\sum_{i=1,2,3}B_i \epsilon_i v_i + |\boldsymbol{\epsilon}|^2v_u\,,\\
  t_i^{(0)}&=D v_i + \epsilon_i(-\mu v_d + B_i v_u +\boldsymbol{\epsilon}\cdot\boldsymbol{v})
  + m_{L_i}^2 v_i\,,\nonumber
\end{align}
where $\boldsymbol{v}^T=(v_1,v_2,v_3)$ and $D=g_Z^2(v_d^2-v_u^2+\sum_i
v_i^2)/8$. The minimization of the potential, $V_\text{linear}=0$,
requires the tadpoles to vanish. Thus,
the vevs can be determined from the system of equations in
(\ref{eq:tree-lvel-tadpoles}) by imposing $t_a=0$ ($a=1,\dots,5$).  In
particular, considering only leading order BRpV terms, the
sneutrino vevs can be written as
\begin{align}
  \label{eq:sneutrino-epsilon-less-basis}
  v_i\simeq \frac{\epsilon_i\,v}{m_{\tilde\nu_i}^{2}}
  (\mu\,c_\beta - B_i \,s_\beta)\,,
\end{align}
where $m_{\tilde\nu_i}^{2}=m_{L_i}^2+M_Z^2(c_\beta^2-s_\beta^2)/2$
(the tree level sneutrino mass) and $t_\beta\equiv \tan\beta=v_u/v_d$.
%
\section{R-parity violating mixings}
\label{sec:sneutrino-mix-analyR}
Without conserved R-parity, there are no quantum numbers to distinguish the
leptons and sleptons from the gauginos and Higgsinos and Higgs bosons respectively.
Therefore, in BRpV several mixings between supersymmetric and non
supersymmetric particles exist: ($i$) neutralinos mix with neutrinos,
($ii$) charginos mix with charged leptons, ($iii$)
Higgs bosons mix with the sneutrinos and
($iv$) charged Higss bosons mix with charged sleptons. These mixings are important for calculating LSP decays, especially mixings of type ($i$) as they allow
to fix---via experimental neutrino data---the size of the BRpV parameters. Since mixings of type ($iv$) are not of interest for
sneutrino decays, in what follows we will only discuss analytical
approximations for mixings of type ($i$)-($iii$).
%
\subsection{Neutralino-neutrino mixings}
\label{sec:neutralino-neu-mix}
In the basis $(\psi^0)^T=(-i\lambda, -i\lambda_3,\tilde H^0_d,\tilde
H^0_u,\nu_i)$ the neutral fermion mass matrix can be written as
\begin{equation}
  \label{eq:Lag-neutral-fermion-mm}
  {\cal  L}_{\psi^0}=-\frac{1}{2}(\psi^0)^T\boldsymbol{M_N}\psi^0 + \mbox{H.c.}
\end{equation}
with
\begin{equation}
  \label{eq:nfmm}
  \boldsymbol{M_N}=
  \begin{pmatrix}
    \boldsymbol{M_{\chi^0}}|_{4\times 4}    & \boldsymbol{M_{\chi\nu}}^T|_{4\times 3}\\
    \boldsymbol{M_{\chi\nu}}|_{3\times 4} & \boldsymbol{0}|_{3\times 3}
  \end{pmatrix}\,,
\end{equation}
where $\boldsymbol{M_\chi}$ is the neutralino mass matrix:
\begin{equation}
	\boldsymbol{M_{\chi^0}} =
	\begin{pmatrix}
		M_1
		&
		0
		&
		-g' v_d/ 2
		&
		g' v_u/2
		\\
		0
		&
		M_2
		&
		g v_d/2
		&
		-g v_u/2
		\\
		-g' v_d/2
		&
		g v_d/2
		&
		0
		&
		-\mu
		\\
		g' v_u/2
		&
		-g v_u /2
		&
		-\mu
		&
		0
	\end{pmatrix},
\end{equation}
 and $M_1$ and $M_2$ are the soft masses for the bino and wino respectively.
$\boldsymbol{M_{\chi\nu}}$ is the $4\times 3$ neutralino-neutrino mixing
matrix given by
\begin{equation}
  \label{eq:neutralino-neutrino-mm}
  \boldsymbol{M_{\chi\nu}}=
  \begin{pmatrix}
    -\frac{1}{2}g'v_i & \frac{1}{2}g v_i & 0 & \epsilon_i
  \end{pmatrix}\,.
\end{equation}
In the Weyl mass eigenstate basis, defined as\footnote{The Majorana
  mass eigenstates are defined as
  $\overline{\chi^0}=(\overline{F_i^0}\;F_i^0)$.}
\begin{equation}
  \label{eq:mass-eigens-neutralino-neu}
  F^0=\boldsymbol{N}\psi^0\,,
\end{equation}
the mass matrix becomes
\begin{equation}
  \label{eq:MM-in-the-mass-eigenstate}
  \boldsymbol{\hat M_N}=\boldsymbol{N}^*\;\boldsymbol{M_N}\;\boldsymbol{N}^\dagger\,.
\end{equation}
Due to the smallness of the BRpV parameters, at order $\epsilon_i$,
$\boldsymbol{M_N}$ can be block diagonalized by decomposing the diagonalizing
matrix $\boldsymbol{N}$ as follows \cite{Hirsch:2000ef}:
\begin{equation}
  \label{eq:N-matrix}
  \boldsymbol{N}=\boldsymbol{\mathcal{N}}\;\boldsymbol{\Xi}\simeq
  \begin{pmatrix}
    \boldsymbol{N_{C}} & \boldsymbol{0}\\
    \boldsymbol{0}       &  \boldsymbol{U_\ell}^\dagger
  \end{pmatrix}
  \begin{pmatrix}
    \mathbb{I}         & \boldsymbol{\xi}^T\\
    -\boldsymbol{\xi}^*       &  \mathbb{I}
  \end{pmatrix}
  =
  \begin{pmatrix}
    \boldsymbol{N_C}                     & \boldsymbol{N_C}\boldsymbol{\xi}^T\\
    -\boldsymbol{U_\ell}^T\;\boldsymbol{\xi}    & \boldsymbol{U_\ell}^T
  \end{pmatrix}
  \,.
\end{equation}
The matrix $\boldsymbol{\Xi}$ block diagonalizes $\boldsymbol{M_N}$ to the form
$\mbox{diag}(\boldsymbol{M_{\chi^0}},\boldsymbol{m_\nu^\text{eff}})$, where
$\boldsymbol{m_\nu}^{\text{eff}}$ is the tree-level light neutrino mass
matrix. The mixing parameters $\xi_{ij}$ can thus be determined to be
\begin{align}
  \label{eq:neutralino-neu-mixing-parameters}
  \xi_{i1}&=\frac{g'M_2\mu}{2|\boldsymbol{M_{\chi^0}}|}\Lambda_i\,,
  &\xi_{i2}&=-\frac{gM_1\mu}{2|\boldsymbol{M_{\chi^0}}|}\Lambda_i\nonumber\\
  \xi_{i3}&=-\frac{\epsilon_i}{\mu}
  + \frac{(g^2M_1+g^{\prime2}M_2)v_u}{4|\boldsymbol{M_{\chi^0}}|}
  \Lambda_i\,,
  &\xi_{i4}&=-\frac{(g^2M_1+g^{\prime2}M_2)v_u}{4|\boldsymbol{M_{\chi^0}}|}
  \Lambda_i\,,
\end{align}
where $|\boldsymbol{M_{\chi^0}}|=-(M_1M_2\mu^2 - 2M_1\mu M_W^2\;c_\beta
s_\beta - 2M_2\mu M_W^2\;c_\beta s_\beta\;t_{\theta_W})$ (with
$t_W=\tan\theta_W$, $\theta_W$ being the weak mixing angle) and
\begin{equation}
  \label{eq:alignment-vector}
  \Lambda_i=\mu v_i + v_d\epsilon_i\,.
\end{equation}
Finally the block diagonal mixing matrices $\boldsymbol{N_C}$ and
$\boldsymbol{U_\ell}$ in (\ref{eq:N-matrix}) diagonlize the neutralino and
neutrino effective mass matrix, which reads
\begin{equation}
  \label{eq:eff-nmm}
  \left(\boldsymbol{m_\nu}^\text{eff}\right)_{ij}
  =\frac{M_1g^2+M_2g^{\prime 2}}{4|\boldsymbol{M_{\chi^0}}|}
  \Lambda_i\Lambda_j\,.
\end{equation}
Since this matrix has two vanishing eigenvalues it can be diagonalized
by only two rotation matrices, namely
\begin{equation}
  \label{eq:leptonic-mix-mat}
  \boldsymbol{U_\ell}=\boldsymbol{U_\ell}(\theta_{23}^\text{BRpV})\;
  \boldsymbol{U_\ell}(\theta_{13}^\text{BRpV})\,,
\end{equation}
with
\begin{equation}
  \label{eq:mixing-angles}
  \tan^2\theta_{23}^\text{BRpV}=\frac{\Lambda_2^2}{\Lambda_3^2}\,,
  \qquad
  \tan^2\theta_{13}^\text{BRpV}=\frac{\Lambda_1^2}{\Lambda_2^2+\Lambda_3^2}\,.
\end{equation}
For sneutrino decays the relevant part of the $\chi-\nu$ mixing turns
out to be the $\boldsymbol{U_\ell}^T\;\boldsymbol{\xi}$ block, that from
eqs. (\ref{eq:neutralino-neu-mixing-parameters}) and
(\ref{eq:leptonic-mix-mat}) can be written as \cite{Diaz:2003as}
\begin{equation}
  \label{eq:neutralino-neutrino-mix-relevantP}
  \boldsymbol{U_\ell}^T\;\boldsymbol{\xi}=
  \begin{pmatrix}
    0 & 0 & -\bar \epsilon_1/\mu & 0\\
    0 & 0 & -\bar \epsilon_2/\mu & 0\\
    a_1|\boldsymbol{\Lambda}| & a_2|\boldsymbol{\Lambda}| 
    & -\bar \epsilon_3/\mu & a_4|\boldsymbol{\Lambda}|  
  \end{pmatrix}
\end{equation}
where $\bar\epsilon_{1,2}=(\boldsymbol{U_\ell}^T)_{(1,2)j}\epsilon_j$ and
$\bar \epsilon_3=\tilde \epsilon_3-a_3|\boldsymbol{\Lambda}|\mu$ with
$\tilde \epsilon_3=(\boldsymbol{U_\ell}^T)_{3j}\epsilon_j$. The coefficients
$a_i$ are given by
\begin{equation}
  \label{eq:ai-coefficients}
  a_1=\frac{g'M_2\mu}{2|\boldsymbol{M_{\chi^0}}|}\,,\;
  a_2=-\frac{gM_1\mu}{2|\boldsymbol{M_{\chi^0}}|}\,,\;
  a_3=\frac{M_{\tilde\gamma}\;v\;s_\beta}{4|\boldsymbol{M_{\chi^0}}|}\,,\;
  a_4=-\frac{M_{\tilde\gamma}\;v\;c_\beta}{4|\boldsymbol{M_{\chi^0}}|}\,,
\end{equation}
with $M_{\tilde\gamma}=g^2M_1+g^{\prime 2}M_2$. Taking into account
eqs. (\ref{eq:leptonic-mix-mat}) and (\ref{eq:mixing-angles}),
explicitly $\bar \epsilon_i$ ($i=1,2$) and $\tilde \epsilon_3$ are
given by
\begin{align}
  \label{eq:epsilon-tilde-parameters}
  \bar \epsilon_1&=\frac{\epsilon_1(\Lambda^2_2 +\Lambda^2_3) 
    - \Lambda_1(\Lambda_2\epsilon_2+\Lambda_3\epsilon_3)}
  {|\boldsymbol{\Lambda}|\,\sqrt{\Lambda^2_2 +\Lambda^2_3}}\,,
  \nonumber\\
  \bar \epsilon_2&=\frac{\Lambda_3\epsilon_2 - \Lambda_2\epsilon_3}
  {\sqrt{\Lambda^2_2 +\Lambda^2_3}}\,,
  \nonumber\\
  \tilde \epsilon_3&=\frac{\boldsymbol{\Lambda}\cdot\boldsymbol{\epsilon}}{|\boldsymbol{\Lambda}|}\,.
\end{align}
%
\subsection{Chargino-charged lepton mass matrices and mixings}
%
\label{sec:cp-even-mm}
In the bases $(\psi^\pm)^T=(-i\lambda^\pm,H_{u,d}^\pm,e_{R,L}^\pm)$ the
chargino charged lepton mass matrix is determined by the following
Lagrangian
\begin{equation}
  \label{eq:chargino-charged-lepton-Lag}
  -{\cal L}_{\psi^\pm}=\frac{1}{2}
  \Psi^T
  \begin{pmatrix}
    \boldsymbol{0}   & \boldsymbol{M_C}^T\\
    \boldsymbol{M_C} & \boldsymbol{0}
  \end{pmatrix}
  \Psi
  +
  \mbox{H.c.}\,,
\end{equation}
where $\Psi^T=(\psi^+,\psi^-)^T$ and in the basis in which the charged
lepton mass matrix is diagonal $\boldsymbol{M_C}$ can be written as
\begin{equation}
  \label{eq:chargino-charged-mm-block-diag}
  \boldsymbol{M_C}=
  \begin{pmatrix}
    \boldsymbol{M_\chi}|_{2\times 2} & \boldsymbol{M_{R\chi}}|_{2\times 3}\\
    \boldsymbol{M_{L\chi}}|_{3\times 2} & \boldsymbol{\hat M_\ell}|_{3\times 3}
  \end{pmatrix}\,.
\end{equation}
The block diagonal matrices correspond to the MSSM chargino and
charged lepton mass matrices whereas the off-diagonal mass matrices
read
\begin{align}
  \label{eq:off-diagonal-mm-chargino-lepton}
  \boldsymbol{M_{R\chi}}=
  \begin{pmatrix}
    0\\
    -\frac{1}{\sqrt{2}}h^E_iv_i
  \end{pmatrix}\,,
  \qquad
  \boldsymbol{M_{L\chi}}=
  \begin{pmatrix}
    \frac{1}{\sqrt{2}}gv_i & -\epsilon_i
  \end{pmatrix}\,.
\end{align}
Defining the Weyl mass eigenstates as\footnote{The corresponding Dirac
  eigenstates are defined as $\overline{\chi^-_i}=(\overline{F^-_i}
  \;F^+_i)$.}
\begin{equation}
  \label{eq:mass-eigenstates-chargino-lepton}
  F^-=\boldsymbol{U}\psi^-
  \quad\mbox{and}\quad
  F^+=\boldsymbol{V}\psi^+\,,
\end{equation}
the diagonal mass matrix $\boldsymbol{\hat M_C}$ is obtained through the
biunitary transformation:
\begin{equation}
  \label{eq:biunitary-trans}
  \boldsymbol{U}\;\boldsymbol{M_C}\;\boldsymbol{V}^T\,.
\end{equation}
Approximate analytical expressions for the mixing matrices $\boldsymbol{U},\boldsymbol{V}$
have been discussed in
\cite{Diaz:2003as,Nowakowski:1995dx,Hirsch:1998kc}.  Here we describe the method for
completeness.

The off-diagonal block matrix
$\boldsymbol{M_{R\chi}}$, being proportional to the charged lepton Yukawa
couplings, can be neglected, and due to the smallness of the BRpV
parameters the mixing matrices $\boldsymbol{U},\boldsymbol{V}$ can be written according to
\begin{align}
  \label{eq:UandV}
  \boldsymbol{U}&=\boldsymbol{\mathcal{U}}\;\boldsymbol{\Xi_L}\simeq
  \begin{pmatrix}
    \boldsymbol{U_L} & \boldsymbol{0}\nonumber\\
    \boldsymbol{0}   & \mathbb{I}
  \end{pmatrix}
  \begin{pmatrix}
    \mathbb{I} & \boldsymbol{\xi_L}^\dagger\\
    -\boldsymbol{\xi_L} & \mathbb{I}
  \end{pmatrix}\,,
  \\
  \boldsymbol{V}&=\boldsymbol{\mathcal{V}}\;\boldsymbol{\Xi_R}\simeq
  \begin{pmatrix}
    \boldsymbol{V_R} & \boldsymbol{0}\\
    \boldsymbol{0}   & \mathbb{I}
  \end{pmatrix}
  \begin{pmatrix}
    \mathbb{I} & \boldsymbol{\xi_R}^T\\
    -\boldsymbol{\xi_R}^* & \mathbb{I}
  \end{pmatrix}\,,
\end{align}
where, in first approximation in the BRpV parameters, the matrices
$\boldsymbol{\Xi}_{\boldsymbol{L},\boldsymbol{R}}$ block-diagonalize the mass matrix, namely
\begin{equation}
  \label{eq:block-diag-Mc}
    \begin{pmatrix}
    \mathbb{I} & \boldsymbol{\xi_L}^\dagger\\
    -\boldsymbol{\xi_L} & \mathbb{I}
  \end{pmatrix}
  \;
  \begin{pmatrix}
    \boldsymbol{M_\chi} & \boldsymbol{0}\\
    \boldsymbol{M_{L\chi}} & \boldsymbol{\hat M_\ell}
  \end{pmatrix}
  \;
  \begin{pmatrix}
    \mathbb{I} & \boldsymbol{\xi_R}^T\\
    -\boldsymbol{\xi_R}^* & \mathbb{I}
  \end{pmatrix}
  \simeq
  \begin{pmatrix}
    \boldsymbol{M_\chi} & \boldsymbol{0}\\
    \boldsymbol{0} & \boldsymbol{\hat M_\ell}
  \end{pmatrix}\,,
\end{equation}
and the matrices $\boldsymbol{U_L}, \boldsymbol{V_R}$ diagonalize in turn the
chargino mass matrix, with the rotation angles given by
\begin{equation}
  \label{eq:chargino-rot-angles}
  \tan2\theta_L=
  -\frac{2\sqrt{2}M_W(M_2 c_\beta + \mu s_\beta)}{M^2_2-\mu^2-2M_W^2c_{2\beta}}
  \,,\quad
  \tan2\theta_R=
  -\frac{2\sqrt{2}M_W(M_2 s_\beta + \mu c_\beta)}{M^2_2-\mu^2-2M_W^2c_{2\beta}}
  \,.
\end{equation}
From equation (\ref{eq:block-diag-Mc}) the matrices $\boldsymbol{\xi}_{\boldsymbol{L},\boldsymbol{R}}$
are found to be
\begin{equation}
  \label{eq:xiR-xiL}
  \boldsymbol{\xi_R}^T=\boldsymbol{M_\chi}^{-1}\;\boldsymbol{\xi_L}^T\;\boldsymbol{\hat M_\ell}
  \quad\mbox{and}\quad
  \boldsymbol{\xi_L}^*=\boldsymbol{M_{L\chi}}\;\boldsymbol{M_\chi}^{-1}\,,
\end{equation}
which implies $\boldsymbol{\xi_R}$ is suppressed with respect $\boldsymbol{\xi_L}$
by a factor $m_\ell/m_\text{susy}$ and thus can be neglected
($\boldsymbol{\Xi_R}=\mathbb{I}_{5 \times 5}$). Explicitly $\boldsymbol{\xi_L}$ can
be written in terms of the BRpV parameters and the coefficients
entering in the chargino mass matrix:
\begin{align}
  \label{eq:xiL-explicitly}
  \xi_{L_{i1}}&=\frac{g}{\sqrt{2}}\frac{\Lambda_i}{|\boldsymbol{M_\chi}|}
  \quad\mbox{with}\quad |\boldsymbol{M_\chi}|=M_2\mu-M_W^2s_{2\beta}\,,\nonumber\\
  \xi_{L_{i2}}&=-\frac{2M_W^2 s_\beta}{v|\boldsymbol{M_\chi}|\mu}\Lambda_i
  -\frac{\epsilon_i}{\mu}\,.
\end{align}
%
\subsection{CP-even neutral scalars mass matrices and mixings}
%
\label{sec:cp-even-mm}
In the basis $(S^{0})^T=(\sigma_d^0,\sigma_u^0,\tilde \nu_i^R)^T$ the
mass matrix of the CP-even neutral scalars $S^0$ is determined by the
following quadratic terms
\begin{equation}
  \label{eq:CP-even-mm}
  {\cal  L}_S^0=\frac{1}{2}(S^0)^T
  \boldsymbol{M_{S^0}}^2
  S^0=\frac{1}{2}(S^0)^T
  \begin{pmatrix}
    \boldsymbol{M_{HH}}^2|_{2\times 2}         & \boldsymbol{M_{H\tilde \nu}}^2|_{2\times 3}\\
    \boldsymbol{M_{H\tilde \nu}}^2|_{3\times 2} & 
    \boldsymbol{M_{\tilde \nu\tilde \nu}}^2|_{3\times3}\,.
  \end{pmatrix}
  S^0\,.
\end{equation}

In what follows we will discuss approximate analytical formulas for
the $\sigma_{u,d}^0-\tilde\nu_i^R$ mixing. To our knowledge, this is
the first time that they are explicitly given.  The entries of the
mass matrix in (\ref{eq:CP-even-mm}) involve the parameters $\mu$,
$\tan\beta$, the soft SUSY breaking coefficients $m_{L_i}$ and $B$,
and the R-parity breaking parameters $\epsilon_i$, $v_i$ and $B_i$. We
use the minimization conditions of the scalar potential
($t_{u,d,i}=0$) to remove the parameters $B_i$. In doing so the matrix
$\boldsymbol{M_{HH}}^2$ can be written in terms of the CP-odd neutral
scalar mass $m_{A^0}$, the sneutrino masses $m_{\tilde \nu_i}$, $M_Z$
and the BRpV parameters $\epsilon_i$ and $v_i$, namely
\begin{align}
  \label{eq:HH-entries}
  (\boldsymbol{M_{HH}}^2)_{11}&=
  m_{A^0}^{2} s_\beta^2 
  + M_Z^2 c_\beta^2
  + \frac{\mu}{v_d}\boldsymbol{\epsilon}\cdot \boldsymbol{v},\nonumber\\
  (\boldsymbol{M_{HH}}^2)_{12}&=
  (\boldsymbol{M_{HH}}^2)_{21}=
  -(m_{A^0}^{2}+M_Z^2)c_\beta s_\beta,\\
  (\boldsymbol{M_{HH}}^2)_{22}&=
  m_{A^0}^{2}c_\beta^2+M_Z^2s_\beta^2
  +\sum_i \bar m_{\tilde\nu_i}^2\frac{v_i^2}{v_u^2}
  -\frac{\mu c_\beta^2}{v_d s_\beta^2}\sum_{i=1,2,3}\epsilon_i^2 v_i^2
  +2\frac{|\boldsymbol{\epsilon}\cdot \boldsymbol{v}|^2}{v_u^2}
  +\frac{1}{4}\frac{g_Z^2}{v_u^2}
  \sum_{\substack{i<j\\ j=1,2,3}}
  v_i^2v_j^2\,.\nonumber
\end{align}
Where the following relations have been used
\begin{align}
  \label{eq:definitions-cp-even}
  m_{A^0}^{2}=\frac{2 B\mu}{s_{2\beta}}\,,\qquad
  \bar m_{\tilde\nu_i}^2=m_{\tilde\nu_i}^{2}+\epsilon_i^2 + \frac{1}{8}g_Z^2v_i^2\,.
\end{align}
Note that phenomenologically consistency requires the inclusion of the
one-loop correction in the $(\boldsymbol{M_{HH}}^2)_{22}$ entry, which in the limit of no stop mixing is
\begin{equation}
  \label{eq:m22-loop}
  (\boldsymbol{M_{HH}}^2)_{22}^\text{1-loop}=\frac{3 m_t^4}{4\pi^2v_u^2}
  \log\left(\frac{m_{\tilde{t}_1}^2\,m_{\tilde{t}_2}^2}{m_t^4}\right)\,.
\end{equation}
See~\cite{Drees:1991mx} for a more complete expression.

In the absence of BRpV, $\boldsymbol{M_{HH}}^2$ corresponds to the
neutral CP-even Higgs mass matrix of the R-parity conserving MSSM, which can be diagonalized via the
rotation matrix
\begin{equation}
  \label{eq:rotation-matrix-CP-even}
  \boldsymbol{R_\alpha}=
  \begin{pmatrix}
    c_\alpha  & s_\alpha\\
    -s_\alpha & c_\alpha
  \end{pmatrix}
  \quad\mbox{with}\quad
  t_{2\alpha}=\frac{m_A^{2(0)}+M_Z^2}{m_A^{2(0)}-M_Z^2}\;t_{2\beta}\,.
\end{equation}
The elements of the $3\times 3$ right-lower block sneutrino
mass matrix in (\ref{eq:CP-even-mm}) are given by
\begin{align}
  \label{eq:sneutrino-mass-matrixplusRpV}
  (\boldsymbol{M_{\tilde\nu\tilde\nu}}^2)_{(i+2)(i+2)}&=\bar
  m_{\tilde\nu_i}^2+\frac{1}{4}v_i^2\,,\nonumber\\
  (\boldsymbol{M_{\tilde\nu\tilde\nu}}^2)_{(i+2)(j+2)}&=
  \frac{1}{4}g^2_Zv_iv_j+\epsilon_i\epsilon_j\qquad
  (\mbox{with $i<j$ and $i=1,2$}).
\end{align}
Finally, for the $\sigma_{u,d}^0-\tilde\nu_i^R$ mixing $2\times 3$
block we have
\begin{align}
  \label{eq:sigma-tildenu-mix}
  (\boldsymbol{M_{H \tilde\nu}}^2)_{1(i+2)}&=-\mu^2\frac{\epsilon_i}{\mu} 
  + M_Z^2\,c_\beta^2\frac{v_i}{v_d}\,,\nonumber\\
  (\boldsymbol{M_{H \tilde\nu}}^2)_{2(i+2)}&=\frac{\mu^2}{t_\beta}\frac{\epsilon_i}{\mu}
   -(\bar m_{\tilde\nu_i}^2+M_Z^2 s_\beta^2)\frac{1}{t_\beta}\frac{v_i}{v_d}
   -\frac{1}{2}M_Z^2c_\beta^3\frac{v_i}{v_d}\sum_{j\neq i}\frac{v_j^2}{v^2_d}
   -\frac{\mu^2}{t_\beta}\frac{\epsilon_i}{\mu}\sum_{i\neq j}\frac{v_j}{v_d}\frac{\epsilon_j}{\mu}\,.
\end{align}
In the mass eigenstate basis defined as
\begin{equation}
  \label{eq:mass-eigenstate-basis-S0}
  S^{\prime 0}=\boldsymbol{R^{S^0}}S^0\,,
\end{equation}
the Lagrangian in (\ref{eq:CP-even-mm}) becomes
\begin{equation}
  \label{eq:Lag-m-eig-S0}
  {\cal  L}_{S^0}=\frac{1}{2}(S^{\prime 0})^T\,\boldsymbol{\hat M_{S^0}^2}\,S^{\prime 0}
  \quad\mbox{with}\quad 
  {\boldsymbol{\hat M_{S^0}}}^2=\boldsymbol{R^{S^0}}{\boldsymbol{M_{S^0}}}^2{\boldsymbol{R^{S^0}}}^T\,,
\end{equation}
where $\boldsymbol{\hat M_{S^0}}^2$ is diagonal.
Assuming real parameters $\boldsymbol{R^{S^0}}$ can be parameterized as
\begin{equation}
  \label{eq:rotS0}
  (\boldsymbol{R^{S^0}})^T=\prod_{\substack{i<j\\ j=1,\cdots,5}}(\boldsymbol{R}_{ij})^T\,,
\end{equation}
where the $\boldsymbol{R}_{ij}\equiv\boldsymbol{R}(\theta_{ij})$ are $5\times 5$
rotation matrices.  

If the $\sigma_{d,u}^0-\tilde\nu_i^R$ mixing is small---as expected
due to the smallness of the BRpV parameters required by neutrino
data---a perturbative diagonalization of the mass matrix in
(\ref{eq:CP-even-mm}) can be done. By neglecting the BRpV parameters
in $\boldsymbol{M_{HH}}^2$ and $\boldsymbol{M_{\tilde\nu\tilde\nu}}^2$ the rotation matrix reduces to
\begin{equation}
  \label{eq:rot-mat-approx-S0}
  \boldsymbol{R^{S^0}}=\boldsymbol{R_{25}}\,\boldsymbol{R_{24}}\,\boldsymbol{R_{23}}\,\boldsymbol{R_{15}}
  \,\boldsymbol{R_{14}}\,\boldsymbol{R_{13}}\,\boldsymbol{R_{12}}\,.
\end{equation}
When acting on $\boldsymbol{M_{S^0}}^2$ the matrix $\boldsymbol{R_{12}}$
diagonalizes the $2\times 2$ block $\boldsymbol{M_{HH}}^2$ according to
$\boldsymbol{\hat M_{HH}}^2=\mbox{diag}(m_{H^0}^2,m_{h^0}^2)$ (where $h^0$
and $H^0$ are the light and heavy CP-even Higgs bosons) and modifies
the BRpV mixing matrix $\boldsymbol{M_{H\tilde\nu}}^2$:
\begin{equation}
  \label{eq:BRpV-mm-modified}
  \boldsymbol{M_{H\tilde\nu}}^2\to
  \begin{pmatrix}
    s_\alpha\,(\boldsymbol{M_{H\tilde\nu}}^2)_{2(i+2)} + c_\alpha (\boldsymbol{M_{H\tilde\nu}}^2)_{1(i+2)}\\
    c_\alpha\,(\boldsymbol{M_{H\tilde\nu}}^2)_{2(i+2)} - s_\alpha (\boldsymbol{M_{H\tilde\nu}}^2)_{1(i+2)}
  \end{pmatrix}\,.
\end{equation}
For the mixing angle we have $\theta_{12}=\alpha$.  The matrices
$\boldsymbol{R_{1(i+2)}}$ eliminate the first row entries in (\ref{eq:BRpV-mm-modified})
and up to ${\cal O}(\epsilon_i^2)$ leave the
block-diagonal matrices $\boldsymbol{\hat{M_{HH}}}^2$ and $\boldsymbol{\hat{M_{\tilde\nu\tilde\nu}}}^2$ diagonal. The rotation angles are given
by
\begin{equation}
  \label{eq:theta-1k-S0}
  \theta_{1(i+2)}\simeq\frac{s_\alpha\,(\boldsymbol{M_{H\tilde\nu}}^2)_{2(i+2)} 
    + c_\alpha (\boldsymbol{M_{H\tilde\nu}}^2)_{1(i+2)}}{m_{H^0}^2 - m_{\tilde\nu_i}^2}\,.
\end{equation}
The matrices $\boldsymbol{R_{2(i+2)}}$ instead eliminate the
second row elements in (\ref{eq:BRpV-mm-modified}) leaving again, up to
order $\epsilon^2_i$, the block-diagonal matrices diagonal. The
rotation angles in this case read
\begin{equation}
  \label{eq:theta-2k-S0}
  \theta_{2(i+2)}\simeq\frac{c_\alpha\,({\boldsymbol{M_{H\tilde\nu}}}^2)_{2(i+2)} 
    - s_\alpha ({\boldsymbol{M_{H\tilde\nu}}}^2)_{1(i+2)}}{m_{h^0}^2 - m_{\tilde\nu_i}^2}\,.
\end{equation}
With these results at hand and neglecting terms of ${\cal
  O}(\theta^2)$ the full rotation matrix in
(\ref{eq:rot-mat-approx-S0}) can be written as
\begin{equation}
  \label{eq:full-rot-S0-final}
  \boldsymbol{R^{S^0}}\simeq
  \begin{pmatrix}
    \boldsymbol{R_\alpha} & \boldsymbol{R_{\tilde\nu}}\\
    \boldsymbol{R_\sigma} & \mathbb{I}
  \end{pmatrix}\,,
\end{equation}
where $\boldsymbol{R_{\tilde\nu}}$ and $\boldsymbol{R_\sigma}$ account for the
sneutrino and CP-even neutral Higgs components of the mass eigenstates,
and read
\begin{equation}
  \label{Rsneutrino-Rhiggs}
  \boldsymbol{R_{\tilde\nu}}=
  \begin{pmatrix}
    \theta_{1(i+2)}\\
    \theta_{2(i+2)}
  \end{pmatrix}\,,
  \qquad
  \boldsymbol{R_{\sigma}}=
  \begin{pmatrix}
    -c_\alpha\theta_{1(i+2)}+s_\alpha\theta_{2(i+2)}&
    -s_\alpha\theta_{1(i+2)}-c_\alpha\theta_{2(i+2)}
  \end{pmatrix}\,.
\end{equation} 
\section{Sneutrino decays}
\label{sec:sneutrino-decays-analyticalR}
With the results of section~\ref{sec:sneutrino-mix-analyR} we are now
in a position to discuss approximate formulas for sneutrino
decays. From now on we will consider only CP-even sneutrino decays,
$\tilde\nu^R$, and so will drop the superscript $R$.  Possible
tree-level two-body sneutrino final states include fermionic modes
$\tilde\nu_i\to l_j^+ l_k^-$, $\nu_j\nu_k$ and $q_k \bar q_k$;
electroweak gauge bosons modes $\tilde\nu_i\to W^+ W^-$ and $Z^0 Z^0$
and Higgs bosons modes $\tilde\nu_i\to h^0 h^0$, $H^0 H^0$, $A^0A^0$
and $H^+H^-$. 

The goal of this paper and section is not a full study of the parameter space and all decays but rather to show that these heavy states can dominate the traditionally considered decay to $b \bar b$ and to identify the relevant parameter space for this dominance.

\subsection{Relevant Lagrangians}
\label{sec:fermionic-modes}
\begin{figure}
  \centering
  \includegraphics[width=10cm,height=3.5cm]{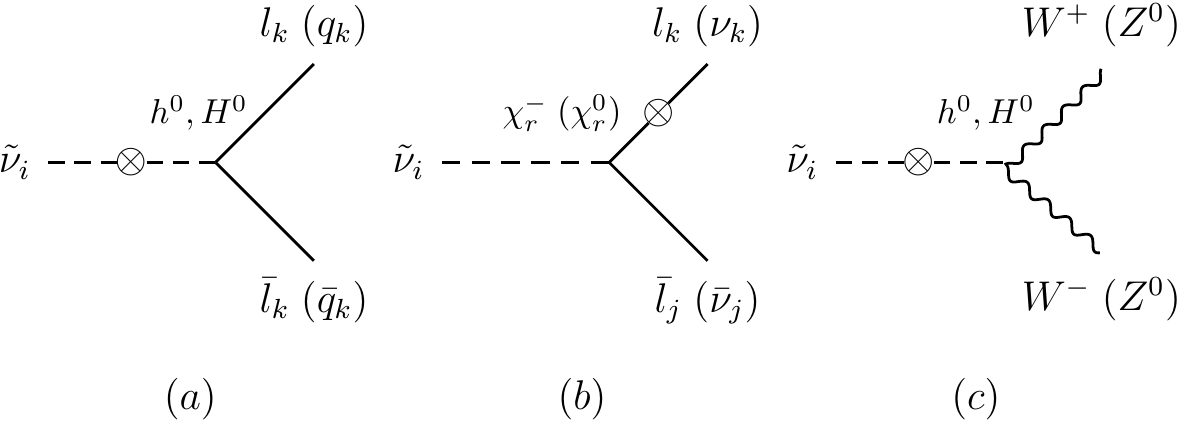}
  \caption{\it BRpV induced sneutrino decay modes. In $(a) \text{ and } (c)$
    sneutrino decays are induced by sneutrino-Higgs mixing while in
    figure~$(b)$ by chargino-charged lepton or neutralino-neutrino
    mixing, depending on whether the final states involves charged
    lepton or neutrinos. The open circles with a cross inside indicate
    a BRpV mixing insertion.}
  \label{fig:sneutrino-decays}
\end{figure}

Taking into account our approximate results for the
chargino-right-handed lepton mixing, $\xi_R\sim
(m_\ell/m_\text{susy})$, mixing with right-handed leptons is zero ($\xi_R\sim 0$, see eq. (\ref{eq:xiR-xiL})), the
full Lagrangian for sneutrino decays into charged leptons is
reduced to~\cite{Hirsch:2000ef}
\begin{align}
  \label{eq:charged-lepton-modes-Lag}
  -{\cal  L}^{(\pm)}= \frac{h^E_k}{\sqrt{2}}\;(\boldsymbol{R^{S^0}})_{(i+2)1}\;
  \overline{l_k^-}\;P_L\;l_k^-
  \tilde\nu_i
  -
  \frac{h^E_i}{\sqrt{2}}\;\boldsymbol{U}_{(j+2)2}\;
  \overline{l_i^-}\;P_L\;l_j^-
  \tilde\nu_i
  +
  \mbox{H.c.}
  \,,
\end{align}
where the mixing matrices are given by eqs. (\ref{eq:xiL-explicitly})
and (\ref{eq:full-rot-S0-final}). Due to the smallness of the BRpV
induced mixing we take $S^{\prime 0}_{(i+2)}\to \tilde\nu_i$. Note
that while the first term in (\ref{eq:charged-lepton-modes-Lag})
necessarily leads to two same-flavor opposite-sign charged lepton
final states the second term can lead to different-flavor signatures.
The corresponding Feynman diagrams for these processes are depicted in
figure~\ref{fig:sneutrino-decays}($(a)$,$(b)$). Being proportional to
SM charged lepton Yukawa couplings these decays are
dominated by final states involving $\tau$'s.

For sneutrino decays into neutrinos (invisible decays $\tilde\nu_i\to\sum_{k,j}\nu_k\nu_j$), the relevant interactions are given by
\begin{equation}
  \label{eq:neutrino-final-states-Lag}
  -{\cal  L}^{(0)}= \frac{1}{2}
  \overline{\nu_k}\left(C_{kji} + C_{jki}\right)\nu_j\;\tilde\nu_i
  +
  \mbox{H.c.}\,,
\end{equation}
where the coupling reads
\begin{align}
  \label{invisible-coupling}
  C_{kji} = \bar N_{k+4}\,\sum_{n=1}^5\,s_n\,\boldsymbol{N}_{(j+4)(n+2)}\,
  (\boldsymbol{R^{S^0}})_{(i+2)n}\,,
\end{align}
where $s_n=(1,-1,1,1,1)$ and
\begin{equation}
  \label{eq:nbar-invisible-modes}
  \bar N_{k+4} = g^\prime\,\boldsymbol{N}_{(k+4)1} - g\,\boldsymbol{N}_{(k+4)2}\,. 
\end{equation}
The elements of the neutral fermion mixing matrix correspond to the
entries of the matrix given in eq. (\ref{eq:N-matrix}). These
interactions induce invisible sneutrino decays as the ones shown in
figure~\ref{fig:sneutrino-decays}$(b)$.

\begin{figure}
  \centering
  \includegraphics[width=8.5cm,height=3.5cm]{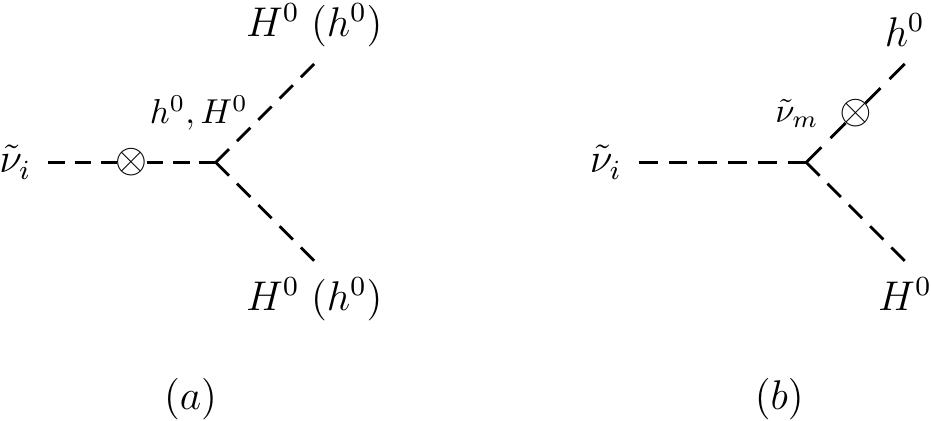}
  \caption{\it BRpV induced sneutrino decays involving Higgs final
    states. The open circles with a cross inside indicate a BRpV
    mixing insertion.}
  \label{fig:sneutrino-decays2}
\end{figure}
In the up and down quark mass eigenstate basis the
sneutrino-quark-quark interactions are dictated by
\begin{equation}
  \label{eq:Lag-senutrino-quarks}
  {\cal  L}^{(q)}=
  -\frac{1}{\sqrt{2}}h^U_k (\boldsymbol{R^{S^0}})_{(i+2)2}\;
  \bar u_k\;u_k\;\tilde\nu_i
  -\frac{1}{\sqrt{2}}h^D_k (\boldsymbol{R^{S^0}})_{(i+2)1}
  \bar d_k\;d_k\;\tilde\nu_i\,.
\end{equation}
As for the charged lepton final states, these decays are controlled by
SM quark Yukawa couplings and thus are dominated by $b\bar
b$ and $t\bar t$, the last one if kinematically allowed.

The Lagrangian for gauge boson final states is given by
\begin{equation}
  \label{eq:Lag-GBFS}
  {\cal  L}^{(V)}=g_V
  \left(
    c_\beta(\boldsymbol{R^{S^0}})_{(i+2)1}
    + s_\beta(\boldsymbol{R^{S^0}})_{(i+2)2} 
    + \frac{v_i}{v} 
  \right)
  \;V^\mu\;V_\mu\;\tilde\nu_i\,,
\end{equation}
with $V_\mu=W_\mu,Z_\mu$ and $g_{W,Z}=g\,M_W, g\,M_Z/c_W$. For Higgs
final states we write the Lagrangian involving $h^0$ and $H^0$\footnote{We 
do not study $\tilde\nu_i\to A^0\,A^0$ and
  $\tilde\nu_i\to H^+\,H^-$ decays and so do not present the Lagrangians
  that govern these interactions.}:
\begin{equation}
  \label{eq:Lagrangian-scalars}
  {\cal  L}^{S^0}= -g_{ijk}\,S^0_j\,S^0_k\,\tilde\nu_i\;,
\end{equation}
where the fully symmetric coupling $g_{(i+2)jk}$ is given by
\begin{equation}
  \label{eq:gijk-coupling}
  g_{ijk}=-\frac{1}{4}\,g_Z^2\,\sum_{n=1}^5\,u_n
  \left[
    (\boldsymbol{R^{S^0}})_{(i+2)n}\,\overline{R}_{jk}
    + (\boldsymbol{R^{S^0}})_{jn}\,\overline{R}_{(i+2)k}
    + (\boldsymbol{R^{S^0}})_{kn}\,\overline{R}_{(i+2)j}
  \right]\,,
\end{equation}
with $u_n=(u_1,u_2,u_3\dots)=(v_d,-v_u,v_1\dots)$ and
\begin{align}
  \label{eq:Rbar-expressions}
  \overline{R}_{jk}=
     (\boldsymbol{R^{S^0}})_{j1}(\boldsymbol{R^{S^0}})_{k1}
    - (\boldsymbol{R^{S^0}})_{j2}(\boldsymbol{R^{S^0}})_{k2}
    + \sum_{n=3}^5(\boldsymbol{R^{S^0}})_{jn}(\boldsymbol{R^{S^0}})_{kn}
    \,.
\end{align}
The interactions in (\ref{eq:Lagrangian-scalars}) for $j=k=1,2$ lead
to decays of type $(a)$ in figure~\ref{fig:sneutrino-decays2} while for
$j=1, k=2$ to those shown in figure~\ref{fig:sneutrino-decays2}$(b)$.
\subsection{Partial decay widths}
\label{sec:higs-modes}
Fermionic final states are dominated by third generation quark and
charged leptons.  Due to the structure of the Higgs-sneutrino mixing,
$\tau\bar\tau$, $b\bar b$ and $t\bar t$ final states are possible
independently of the sneutrino flavor, whereas $\bar\tau (e,\mu)$ final
states are only sizable for tau and mu sneutrinos.  Neglecting the final
state masses, the partial decay widths for $\bar\tau (e,\mu)$,
$\tau\bar\tau$ and $b\bar b$ decays can be written as
\begin{align}
  \label{eq:bbar-ttbar-partial-widths}
  \Gamma(\tilde\nu_i\to \bar l_i l_k)&=
  \frac{m_{l_i}^2\,G_F}{4\sqrt{2}\pi\,c_\beta^2}
  m_{\tilde\nu_i}{\boldsymbol{U}}^2_{(k+2)2}\,,\\
  \label{eq:bbar-ttbar-partial-widths1}
  \Gamma(\tilde\nu_i\to \tau\bar\tau)&=
  \frac{m_\tau^2\,G_F}{4\sqrt{2}\pi\,c_\beta^2}
  m_{\tilde\nu_i}
  \left[
    (\boldsymbol{R^{S^0}})_{(i+2)1} - \boldsymbol{U}_{52}\delta_{i3}
  \right]^2\,,\\
  \label{eq:bbar-ttbar-partial-widths2}
  \Gamma(\tilde\nu_i\to b\bar b)&=
  \frac{3 m_b^2\,G_F}{4\sqrt{2}\pi\,c_\beta^2}
  m_{\tilde\nu_i}(\boldsymbol{R^{S^0}})^2_{(i+2)1}\,,
\end{align}
where $G_F$ is the Fermi constant. For invisible decay modes the
partial decay width, summing over lepton flavors, can be written
according to\footnote{These modes were
  studied in ref.  \cite{de Campos:1995av}}
\begin{equation}
  \label{eq:invisible-decay-width}
  \Gamma(\tilde\nu_i\to \sum_{k,j}\nu_k\nu_j)=
  \frac{m_{\tilde\nu_i}}{16\pi}\,
  \sum_{k,j}\,(C_{kji}+C_{jki})^2\,,
\end{equation}
with $C_{kji}$ given by (\ref{invisible-coupling}). For $t\bar t$
final states the phase space factors are relevant, accordingly the
corresponding decay width reads
\begin{equation}
  \label{eq:tt-partial-width}
  \Gamma(\tilde\nu_i\to t\bar t)=
  \frac{3 m_t^2\,G_F}{4\sqrt{2}\pi\,s_\beta^2}
  m_{\tilde\nu_i}
  (\boldsymbol{R^{S^0}})_{(i+2)2}^2
  \left(1 - 4\frac{m_t^2}{m_{\tilde\nu_i}^2}\right)^{3/2}
\end{equation}

For Gauge boson final states the partial decay width is given by
\begin{equation}
  \label{eq:GB-partial-DW}
    \Gamma(\tilde \nu_i \to VV) = \frac{G_F m_{\tilde \nu_i}^3 }{16
    \sqrt 2 \, \pi} \delta_V \sqrt{1 - 4 \frac{M_V^2}{m_{\tilde
        \nu_i}^2}} \, \left(1 - 4 \frac{M_V^2}{m_{\tilde \nu_i}^2} +
    12 \frac{M_V^4}{m_{\tilde \nu_i}^4}\right) \left| \mathcal
    A_i^V\right|^2\,,
\end{equation}
where $V=Z,W$, $\delta_{Z,W}=1,2$ and the amplitude $A_i^V$ reads
\begin{equation}
  \label{eq:vector-boson-amplitude}
  \mathcal{A}^V_i = c_\beta (\boldsymbol{R^{S^0}})_{(i+2)1} 
  + s_\beta (\boldsymbol{R^{S^0}})_{(i+2)2} + \frac{v_i}{v}\,.
\end{equation}
For $V\,V^*$ final states the partial decay width is given
by \cite{Djouadi:1995gv}
\begin{equation}
  \label{eq:vvstar-differential}
  \frac{d\Gamma(\tilde\nu_i\to VV^*)}{dx_1dx_2}=
  K_{\tilde\nu_i VV}\frac{(1-x_1)(1-x_2)+\kappa_V(2x_1+2x_2-3+2\kappa_V)}
  {(1-x_1-x_2)^2+\kappa_V\gamma_V}\,,
\end{equation}
with
\begin{equation}
  \label{eq:ksnuVV}
  \kappa_V=\frac{M_V^2}{m_{\tilde\nu_i}^2}\,,\quad
  K_{\tilde\nu_i VV}=\frac{3 G_F^2 M_W^4}{16\pi^3}|{\cal A}_V^i|^2
  m_{\tilde\nu_i}\,3\delta'_V
\end{equation}
and
\begin{equation}
  \label{eq:deltaW-deltaZ}
  \delta'_W=1\,,\quad \delta'_Z=\frac{1}{c_W^4}
  \left(\frac{7}{12}-\frac{10}{9}\,s_W^2+\frac{40}{9}\,s_W^4\right)\,.
\end{equation}
The integration variables $x_{1,2}$ lie in the ranges
$x_1=[1-x_2-\kappa_V,1-\kappa_V/(1-x_2)]$ and $x_2=[0,1-\kappa_V]$.
The parameter $\gamma_V=\Gamma_V^2/M_V^2$ ($\Gamma_V$ being the total
decay width of the gauge boson $V$) allows a smooth transition in the
threshold region where the off-shell gauge boson becomes on-shell
($m_{\tilde\nu_i}=2M_V$), the calculation of $\Gamma(\tilde\nu_i\to
VV^*)$ thus requires numerical integration over the variables
$x_{1,2}$ in the transition region. Outside that region, i.e. for
$m_{\tilde\nu_i}\lesssim 2 M_V - \Gamma_V$ it can be written according
to \cite{Djouadi:2005gi}
\begin{equation}
  \label{eq:VV*-final-states}
  \Gamma(\tilde \nu_i \to VV^*) = \frac{3 \, G_F^2 M_V^4 }{16
    \sqrt 2 \, \pi} m_{\tilde \nu_i} \delta'_V R_T \left(
    \frac{M_V^2}{m_{\tilde \nu_i}^2}\right) \left| \mathcal
    A_i^V\right|^2\,,
\end{equation}
where $R_T(x)$ is a kinematic function given by
\begin{align}
  \label{eq:VV*-kinematic-function}
  R_T(x) &= \frac{3(1-8 x + 20 x^2)}{\sqrt{4x -1}} 
  \arccos \left(\frac{3x -1}{2x^{3/2}}\right) 
  - \frac{1-x}{2x} (2-13x+47x^2)\nonumber\\
  &- \frac{3}{2} (1-6x+4x^2) \log x\,.
\end{align}

Finally for CP-even Higgs bosons final states the partial decay width
can be written as
\begin{align}
  \label{eq:HB-partial-DW-general}
  \Gamma(\tilde\nu_i\to S_j^0S_k^0)=\frac{g_{(i+2)jk}^2}{16\pi\,m_{\tilde\nu_i}^3}
  \lambda^{1/2}(m_{\tilde\nu_i}^2,m_{S_j^0}^2,m_{S_k^0}^2)
  \qquad (j,k = 1,2)\,,
\end{align}
where $\lambda(a,b,c)=(a-b-c)^2-4bc$, $S_{1,2}^0=H^0,h^0$ and the
dimensionful coupling $g_{(i+2)jk}$ is given in
(\ref{eq:gijk-coupling}). For the particular case $j=k$ ($h^0h^0$ and
$H^0H^0$ final states) the width reduces to
\begin{equation}
  \label{eq:HB-partial-DW}
  \Gamma(\tilde\nu_i\to S_j^0S_j^0)=\frac{g_{ijj}^2}{32\pi\,m_{\tilde\nu_i}}
  \sqrt{1 - 4\frac{m_{S_j^0}^2}{m_{\tilde\nu_i}^2}}\,.
\end{equation}
\subsection{Constraints on BRpV parameters from neutrino data}
\label{sec:constraints-neu-physics}
Neutrino data plays an important role in BRpV, since BRpV parameters contribute to neutrino masses. Therefore, even if BRpV is not solely responsible for neutrino masses, \textit{e.g.}~\cite{Mohapatra,Buchmuller:2007ui,AristizabalSierra:2003ix,Ghosh:2010hy, Barger:2010iv}, one must not saturate neutrino masses via BRpV. Therefore it would be worthwhile to discuss this correlation briefly here. We proceed by assuming BRpV as the sole source of neutrino masses and later consider relaxing this assumption.

At tree-level, BRpV allows for only one massive neutrino, as mentioned earlier. One-loop contributions then leave only one generation massless. Furthermore, for the approximations made in this paper, the tree-level mass is larger than the one-loop mass. The upshot of all this is that only the so-called normal hierarchy of neutrino masses is allowed. Recent neutrino data~\cite{Tortola:2012te,GonzalezGarcia:2012sz,Fogli:2012ua} then specifies the following neutrino masses:
\begin{equation}
	m_1 = 0, \quad \quad m_2 = \sqrt{\Delta m_{21}^2} = 0.00873 \text{ eV},
	\quad \quad m_3 = \sqrt{\Delta m_{31}^2} = 0.0505 \text{ eV}.
\end{equation}
In terms of the model parameters
\begin{equation}
  \label{eq:m-atmospheric}
  \Delta m_{32}^\text{BRpV}
  =\frac{M_{\tilde\gamma}}{4|\boldsymbol{M_\chi^0}|}
  \,|\boldsymbol{\Lambda}|^2\, .
\end{equation}
The experimental values of the atmospheric and reactor angles yield two
additional constraints given by eq. (\ref{eq:mixing-angles}). Therefore,
the atmospheric sector entirely fixes the $\Lambda_i$ parameters.

\begin{figure}
  \centering
  \includegraphics[width=7.5cm,height=6.5cm]{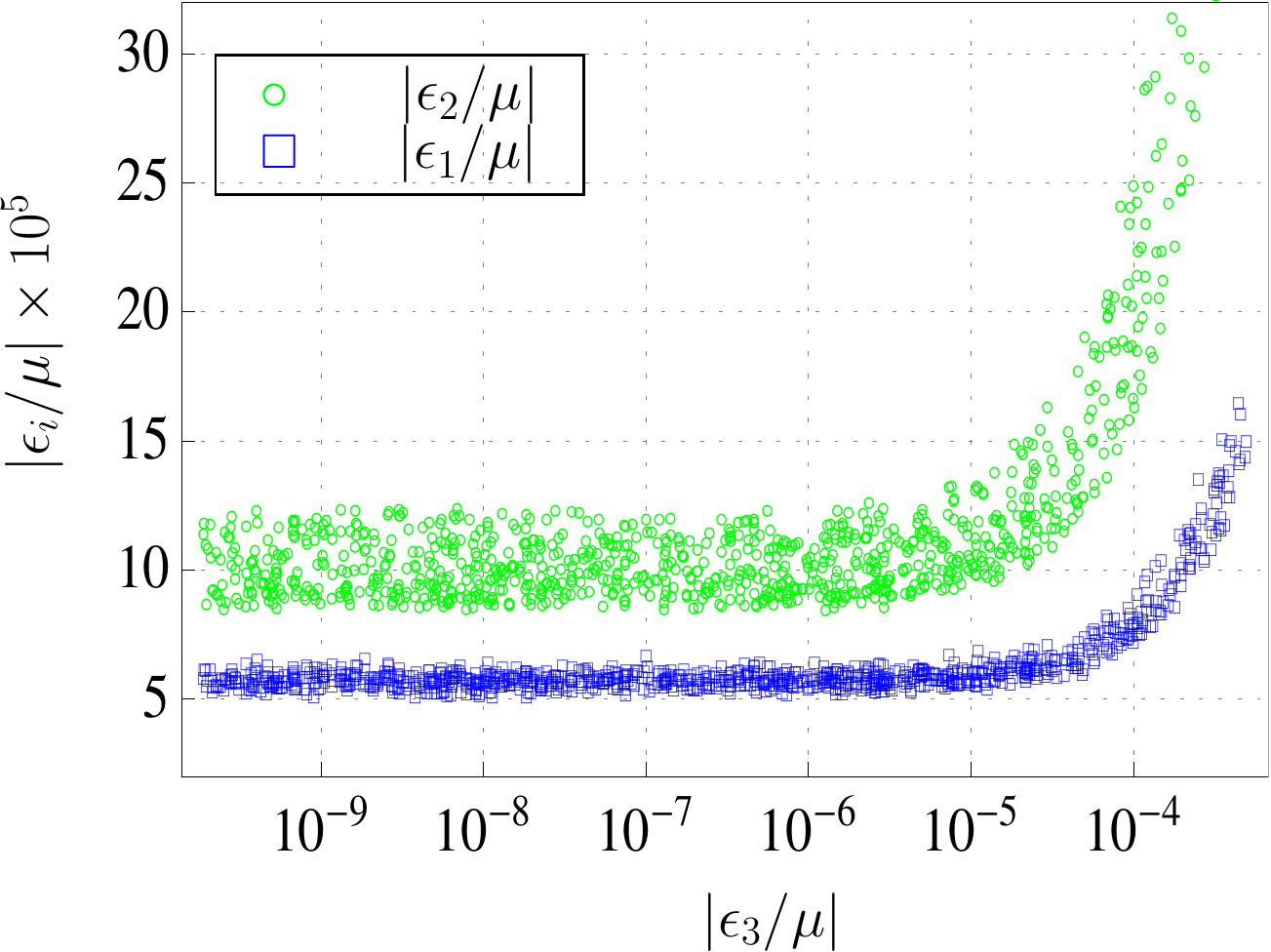}
  \hfill
  \includegraphics[width=7.7cm,height=6.7cm]{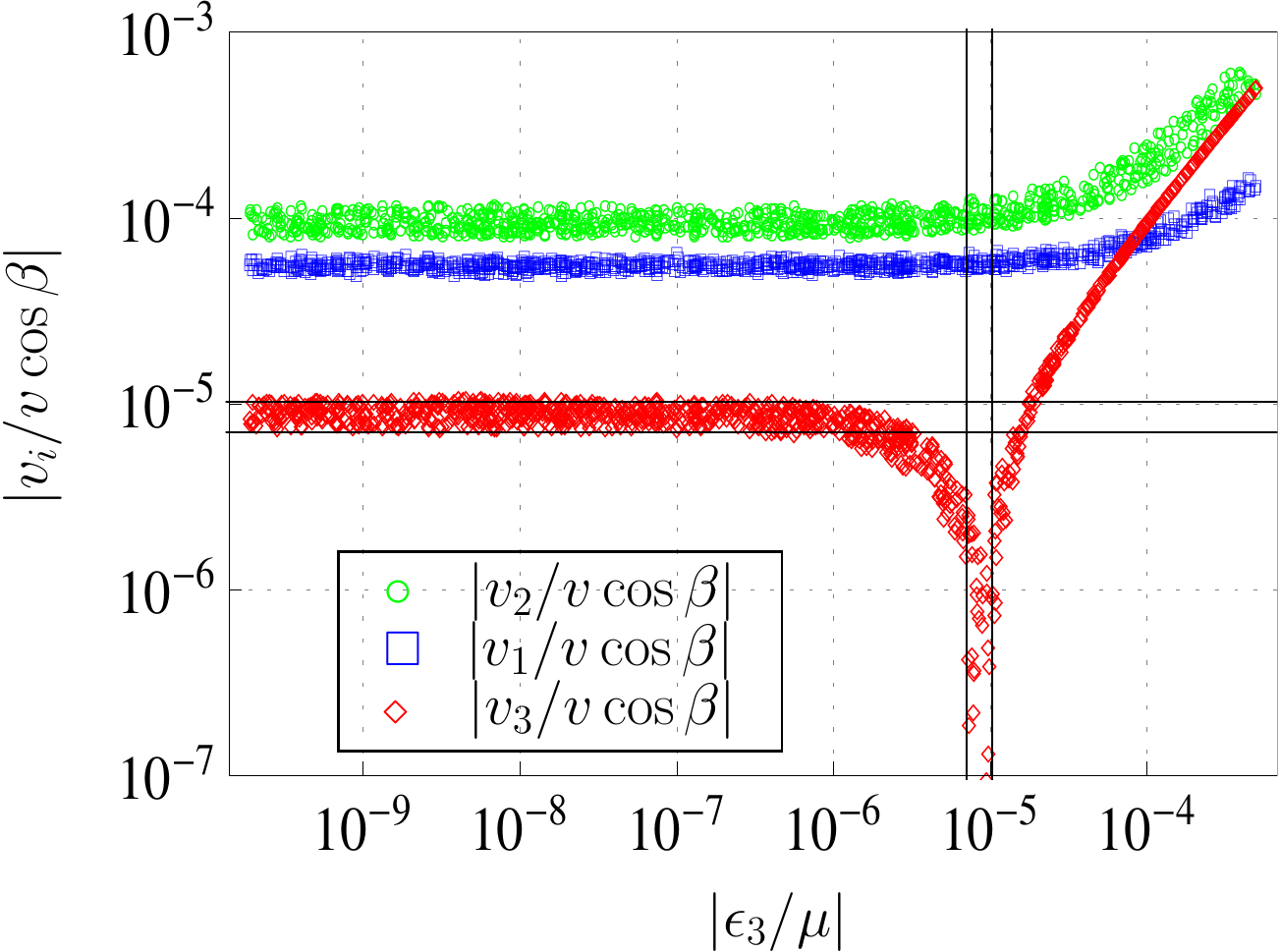}
  \caption{\it Numerical ranges for $\epsilon_i$ ($i=1,2,3$) parameters
    and sneutrino vevs as required for explaining neutrino data for the R-parity conserving parameters discussed in the text. The
    horizontal (vertical) solid lines in the right-panel plot indicate
    the values where $\Lambda_3$ is dominated by $v_3$ ($\epsilon_3$),
    i.e.  where $\Lambda_3\simeq \mu v_3$ ($\Lambda_3\simeq v\cos\beta
    \,\epsilon_3$).}
  \label{fig:epsilons-vevs}
\end{figure}
Constraints on the $\epsilon$ parameters arise from the
solar sector. As long as the
one-loop contribution to neutrino masses is smaller than the
tree-level one (an assumption we use throughout), $\Delta m_{21}^\text{BRpV}$ is accurately determined by
$b-\tilde b$ one-loop diagrams \cite{Diaz:2003as}:
\begin{equation}
  \label{eq:solar-mass-scale}
  \Delta m_{21}^\text{BRpV}=\frac{3}{8\pi^2}\sin2\theta_{\tilde b}
  \frac{m_b^3}{v^2\,c_\beta^2}
  \Delta B_0^{\tilde b_2\tilde b_1}\frac{(\bar\epsilon_1^2 + \bar\epsilon_2^2)}{\mu^2}\,,
\end{equation}
where $\theta_{\tilde{b}}$ stands for sbottom mixing, $\bar
\epsilon_{1,2}$ are defined in
eq. (\ref{eq:epsilon-tilde-parameters}) and $\Delta B_0^{\tilde
  b_2\tilde b_1}=B_0(0,m_{\tilde b_2}^2,m_b^2)-B_0(0,m_{\tilde
  b_1}^2,m_b^2)$ (with $B_0(0,x,y)$ is a scalar Passarino-Veltman
function \cite{Passarino:1978jh}). Note that due to $m_b\ll m_{\tilde b_{1,2}}$, 
\begin{equation}
  \label{eq:DeltaB-approx}
  \Delta B_0^{\tilde b_2\tilde b_1}\simeq 
  \log
  \left(
    \frac{m^2_{\tilde b_2}}{m^2_{\tilde b_1}}
  \right)\,.
\end{equation}
Moreover the solar mixing angle can be written as \cite{Diaz:2003as}
\begin{equation}
  \label{eq:solar-mixing}
  \tan^2\theta_{12}^\text{BRpV}=\frac{\bar\epsilon_1^2}{\bar\epsilon_2^2}\,.
\end{equation}
Thus, 
eqs. (\ref{eq:solar-mass-scale}) and (\ref{eq:solar-mixing}) provide
two constraints and determine, from
eqs.~(\ref{eq:epsilon-tilde-parameters}), $\epsilon_{1,2}$
($\epsilon_{1,3}$ or $\epsilon_{2,3}$) as a function of $\epsilon_3$
($\epsilon_2$ or $\epsilon_1$). Once the $\Lambda$'s and $\epsilon$'s
are fixed the sneutrino vevs are automatically fixed as well through
\begin{equation}
  \label{eq:sneutrino-vevs-Lambdas-epsilons}
  v_i=\frac{\Lambda_i - v_d\epsilon_i}{\mu}\,,
\end{equation}
see eq. (\ref{eq:alignment-vector}). This in turn fixes all the relevant mixings for sneutrino decays
($\chi^0-\nu$, $\chi^--\ell^-_L$, $\sigma_{d,u}^0-\tilde\nu^R_i$) once
the R-parity conserving supersymmetric parameters are
specified. With this knowledge in hand, we start to explore the consequences of the neutrino sector on the BRpV parameters. We note that our results have been verified using SPheno \cite{Porod:2003um}\footnote{We thank
  Werner Porod for fixing a bug in the decay routines which allowed us
  to calculate sneutrino decays to gauge bosons, discussed in
  section~\ref{sec:sneu-tau-phenomenology}}.
  
  Figure~\ref{fig:epsilons-vevs} shows typical values for
$\epsilon_{1,2}$ and sneutrino vevs as a function of $\epsilon_3$. The
plots were obtained by fixing $\theta_{\tilde b}=\pi/16$,
$\tan\beta=10$ and
$$
	(\mu, \ M_1, \ M_2, \ m_{\tilde b_2}, \ m_{\tilde b_1}, \ m_A, \ m_{{\tilde\nu}_i})
=(650,550,600,10^3,700,10^3,300) \text{ GeV},
$$
where we have assumed a conservative lower bound on the sneutrino mass of 100~GeV, see~\cite{Abdallah:2003xe} for more details. We will use this parameter point throughout the paper. Neutrino observable $(\Delta m_{32},\Delta m_{21},\theta_{ij})$ are varied 
 in their
3$\sigma$ experimental range
\cite{Tortola:2012te,GonzalezGarcia:2012sz,Fogli:2012ua}. We further assume that
$\epsilon_3$ 
is positive. For these assumptions we see from the left-hand side that $\epsilon_{1}$ and $\epsilon_{2}$ have a lower bound and from the right-handed side that this lower bound is larger than $v_1$ and $v_2$ respectively. On the other hand, since $\epsilon_3$ is undetermined, $v_3$ dominates it up to about $|\epsilon_3/\mu| = 10^{-5}$. While allowing negative values for the parameters changes this picture, it leaves one important qualitative property the same: \textbf{only in one generation, $j$, can the sneutrino vev, $v_j$, be larger than the bilinear mixing term, $\epsilon_j$, when BRpV is the sole contributor to neutrino masses.} This will have important consequences in the next section.

Also of note in the left-panel plot, $\epsilon_3$ obeys only
an upper bound determined by the condition
$({\boldsymbol{m_\nu}^{\text{eff}}})^{\text{tree}}>({\boldsymbol{m_\nu}^{\text{eff}}})^{\text{1-loop}}$,
which in terms of the bilinear R-parity violating parameters
translates into $|\boldsymbol{\Lambda}|>|\boldsymbol{\epsilon}|^2$. In
contrast the $\epsilon_{1,2}$ parameters, due to solar neutrino
physics constraints, are forced to lie in a ``narrow'' range and are
such that a region where $\epsilon_{1,2}\gg \epsilon_3$
exist. Consequently, while $\Lambda_{1,2}$ are mostly determined by
$\epsilon_{1,2}$, $\Lambda_3$ is controlled by $v_3$ in the region
where $\epsilon_3/\mu\lesssim 10^{-7}$, as demonstrated by the
horizontal solid lines in figure~\ref{fig:epsilons-vevs} (right-panel)
which correspond to $\Lambda_3\simeq \mu v_3$.

\begin{figure}
  \centering
  \includegraphics[width=9cm,height=6.7cm]{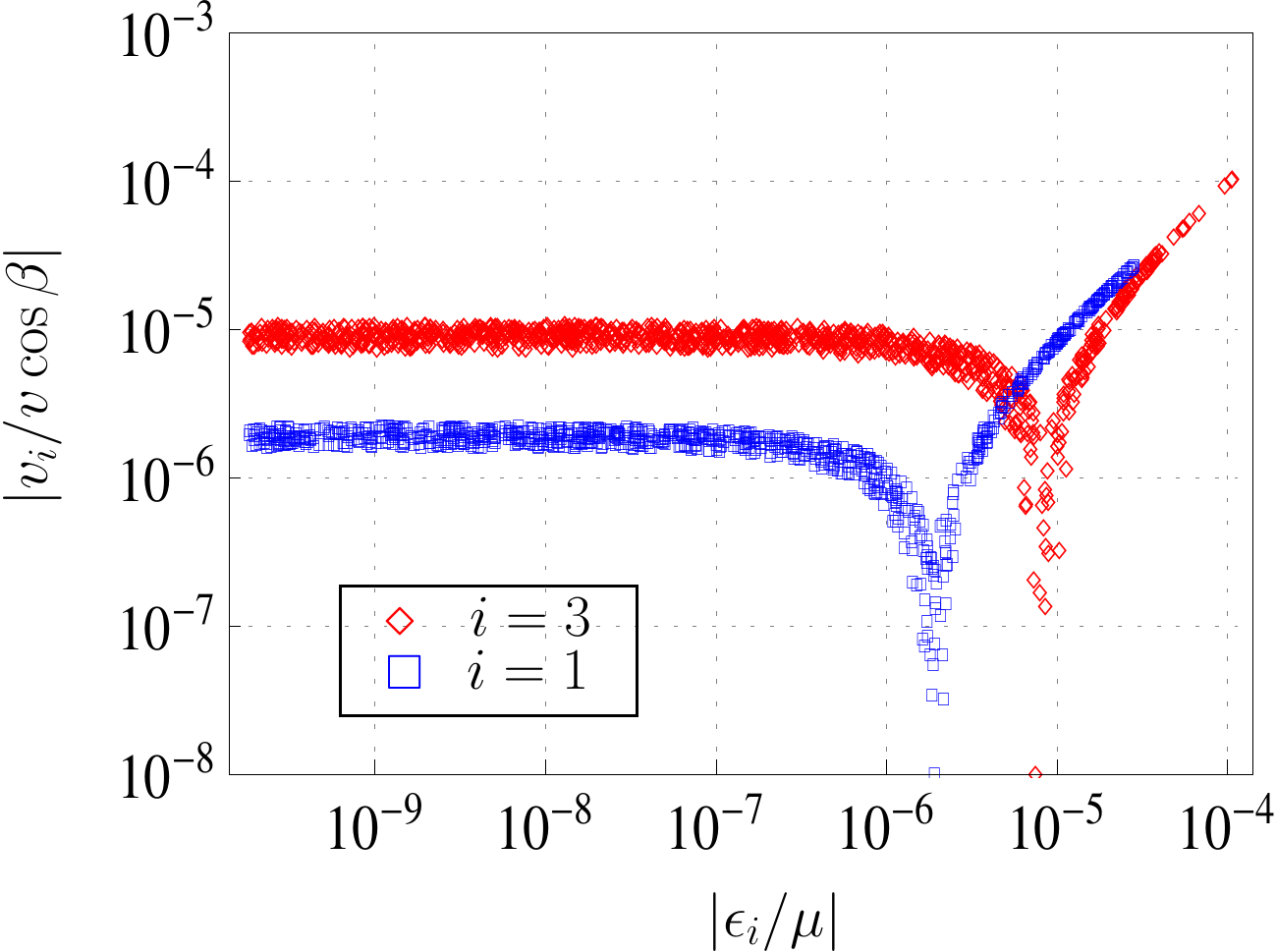}
  \caption{\it Values of R-parity breaking parameters for which BRpV is not the sole source of neutrino physics. The parameters
    are such that $\Delta m_{32}^\text{BRpV}$ fits the experimental
    $3\sigma$ range but $\Delta m_{21}^\text{BRpV}$ falls below the
    measured value (see the text for more details).}
  \label{fig:vevs-noatmospheric-nosolar}
\end{figure}

The bilinear R-parity breaking parameters selected as described above
satisfy neutrino data, and thus lead to BRpV models that can account
for neutrino masses and mixings. However, it might be that these
parameters are not sufficiently large to account for the neutrino mass
scales. In that case their contribution to the atmospheric and solar
masses are still determined by eqs. (\ref{eq:m-atmospheric}) and
(\ref{eq:solar-mass-scale}) but are such that, for example, $\Delta
m_{31}^\text{BRpV}<(\ll)\Delta m_{31}^\text{Exp}$ and $\Delta
m_{21}^\text{BRpV}<(\ll)\Delta m_{21}^\text{Exp}$. Figure
\ref{fig:vevs-noatmospheric-nosolar} shows the results for an
illustrative case where the BRpV model fits the atmospheric mass scale
as well as the atmospheric and reactor angles in their $3\sigma$
experimental range, but the contribution to $\Delta m_{21}$ is
subdominant\footnote{Models of this kind have been considered in the
  literature, see e.g. \cite{Ma:2002xta,AristizabalSierra:2003ix}.}.
Note that we plot only $v_{1,3}$, as we have found that
$v_2\sim v_3$. This result as well as $v_1\ll v_3$ in the region of
small $\epsilon_{1,3}$ are due to the constraints arising from fitting
$\theta_{23}$ and $\theta_{13}$.

In summary, if data is not explained by the BRpV parameters all the
$\epsilon$'s can be small mainly due to the absence of the solar data
constraint, thus implying that, in this case, a region where the three
sneutrino vevs are large ($\Lambda_i\simeq \mu\,v_i$) exist.
%
\subsection{LSP sneutrino phenomenology}
\label{sec:sneu-tau-phenomenology}
We are finally ready to address the main aim of this paper: showing
that sneutrino LSP decays into heavy SM pairs, invisible modes and different
lepton flavor final states \textit{i.e.}  $W^+ W^-, ZZ, h^0 h^0, t
\bar t$; $\nu\nu$ and $l_i^+l_k^-$ ($i\neq k$) can dominate over the
traditionally considered $b\bar b$ mode, which was typically considered as the dominate mode~\cite{Aristizabal
  Sierra:2004cy}.

We begin with a very general study ignoring all neutrino constrains
(SUSY parameters can always be chosen in such a way so that BRpV does
not saturate the neutrino masses) and exploring in the
$\epsilon_i$-$v_i$ parameter space where the aforementioned modes
dominate the $b \bar b$ decays of $\tilde \nu_i$. Without neutrino
constraints, these decay properties are independent of generation. We
scan over the following values:
\begin{align*}
  \tan \beta = 2& - 50, \ \ m_A = 500 - 1000 \text{ GeV}, \ \
  m_{\tilde \nu_i} = 161 - 500\text{ GeV},
  \\
    M_{1,2}=500 - &1000 \text{GeV}, \ \
  v_i = 10^{-8} - 10^{-4} \text{ GeV}, \ \ \epsilon_i = 10^{-8} -
  10^{-4} \text{ GeV},
\end{align*}
\begin{figure}
  \centering
  \includegraphics[width=7.5cm,height=6.5cm]{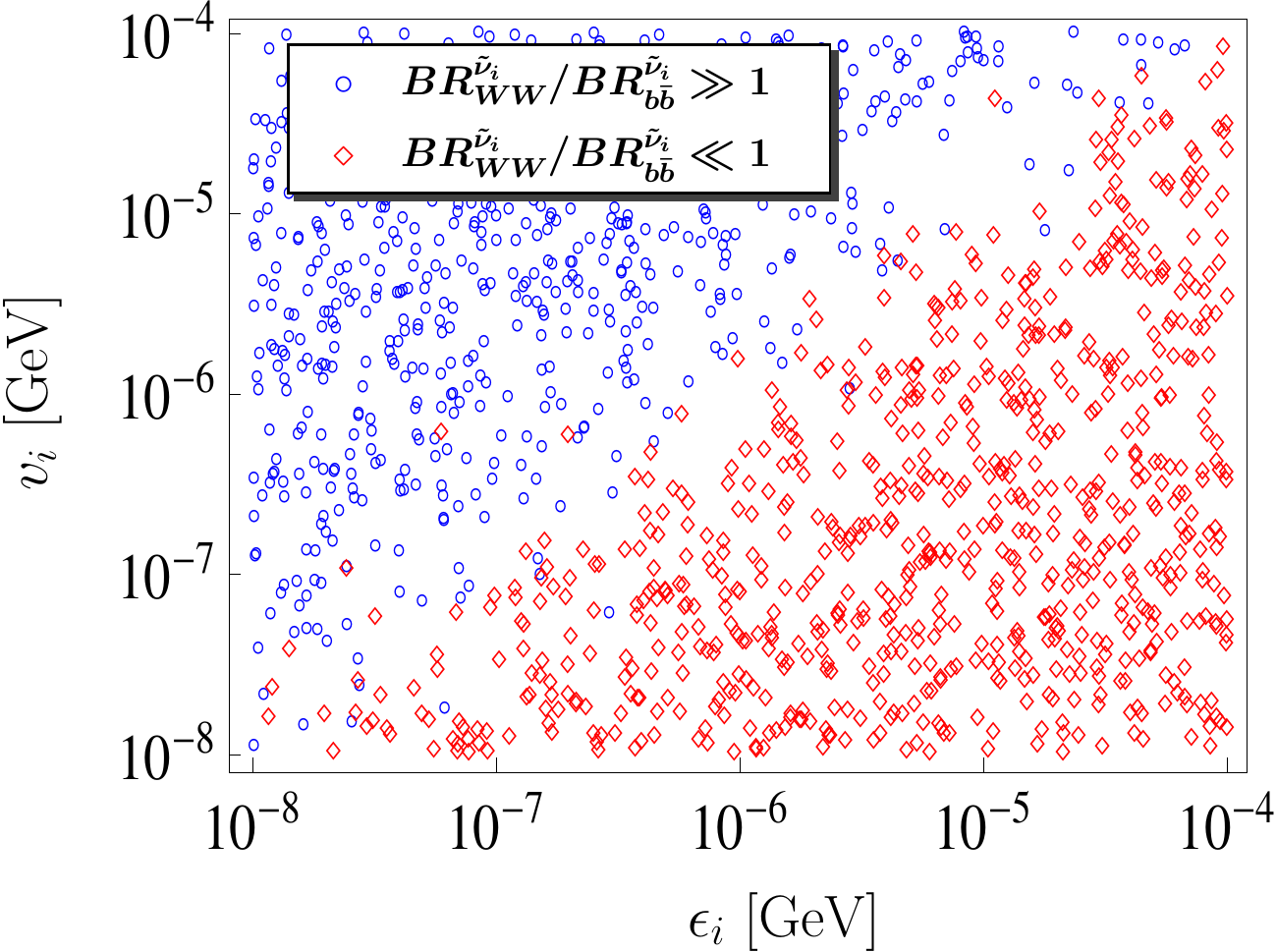}
  \hfill
  \includegraphics[width=7.5cm,height=6.5cm]{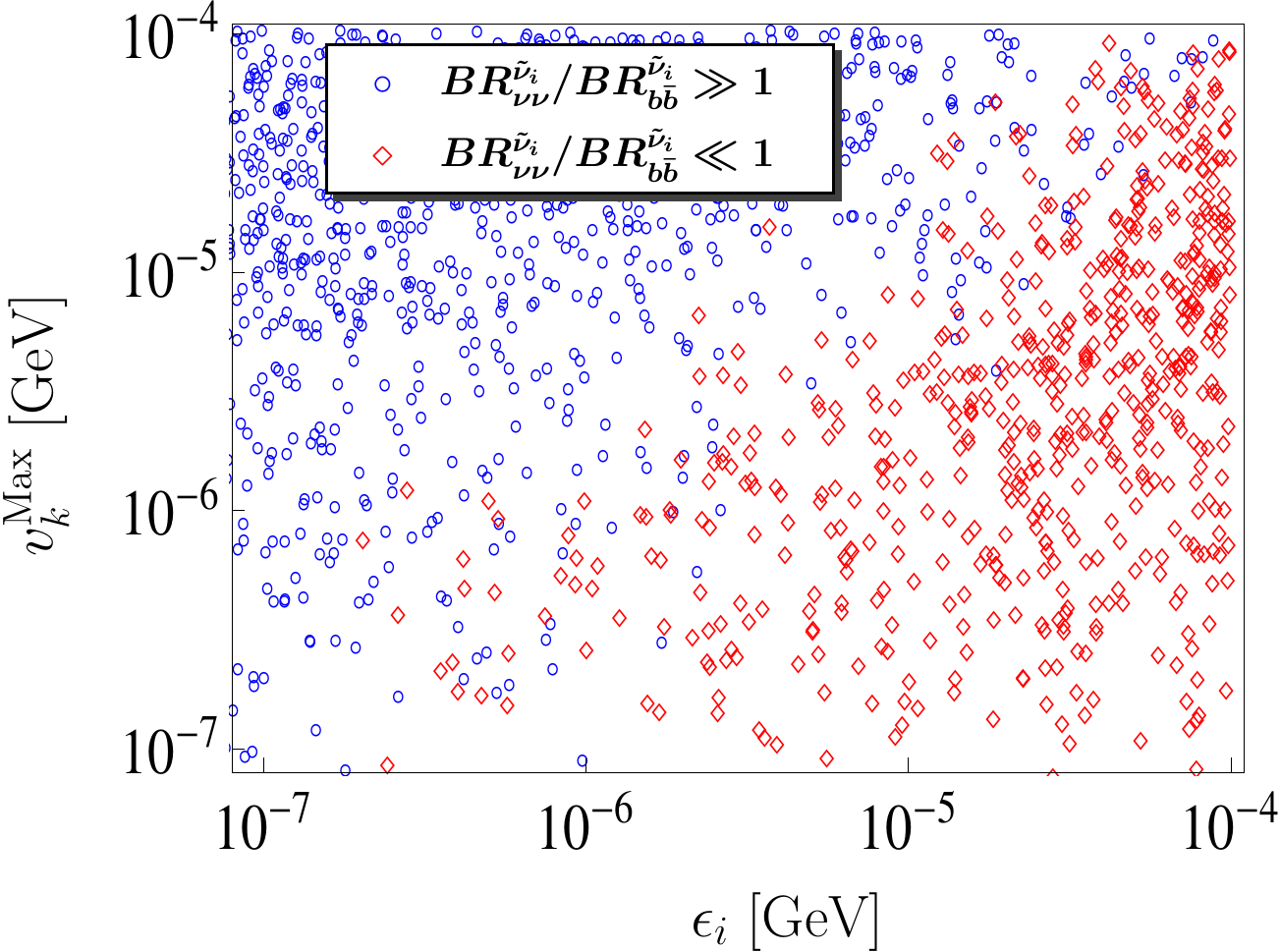}
  \caption{\it Results of a scan over SUSY parameters, as shown in the
    text, comparing the sneutrino decay width into $W^+ W^-$ and
    $\nu\nu$ with the $b\bar b$ final state. It has been assumed that
    BRpV does not saturate neutrino masses. Blue points indicate where
    $\mbox{BR}^{\tilde\nu_i}_{WW}/\mbox{BR}^{\tilde\nu_i}_{b\bar b}\gg
    1$ (left-hand side plot) or
    $\mbox{BR}^{\tilde\nu_i}_{\nu\nu}/\mbox{BR}^{\tilde\nu_i}_{b\bar
      b}\gg 1$ (right-hand side plot) while the red points the
    opposite.  The invisible mode involves a sum over final state
    lepton flavors, and so its importance is determined by the
    relative size between $\epsilon_i$ and any of the three sneutrino
    vevs. The vertical axes in the right-hand side plot corresponds to
    the largest sneutrino vev obtained when randomly scanning the
    parameter space.  These plots therefore show an important general
    result: the $W^+ W^-$ and invisible modes have larger branching
    fractions than the $b\bar b$ mode only when the sneutrino vev is
    much larger than its corresponding bilinear $\epsilon$ parameter.}
  \label{fig:sneu-WW-bb}
\end{figure}
and $\mu$ below 1000~GeV but large enough so that the sneutrino is the
LSP. We also use $m_h =125$~GeV as suggested by recent LHC
results. For $W^+W^-$ and $\nu\nu$ modes the results are displayed in
figure~\ref{fig:sneu-WW-bb} where blue dots indicate the points at
which these channels have partial decay widths much larger than the
$b\bar b$ mode, while the red points show the opposite. It is striking
that the ratio of the partial widths are relatively independent of
R-parity conserving SUSY parameters and depend only on the BRpV
parameters. From this figure, one can conclude that the $W^+ W^-$ and
the $\nu\nu$ partial widths dominate the $b \bar b$ width only when
the sneutrino vev is much larger than the bilinear mixing parameter
and, as will be shown below, this condition holds for the other heavy
SM final states.  Couple this with the main result of the previous
section: satisfying neutrino masses solely through BRpV means that
$v_i \gg \epsilon_i$ can only be satisfied in one generation,
indicates that heavy final states as well as invisible modes can only
have branching fractions larger than the $b\bar b$ mode for one
generation of sneutrinos.

We now turn to the discussion of the different lepton flavor final
states. These modes ($\tilde \nu_i \to l_i^+l_k^-$), being induced by chargino charged
lepton mixing, are mainly controlled by the bilinear mixing parameters for the flavor of the final lepton which differs from the sneutrino flavor (see
eq. (\ref{eq:xiL-explicitly}) and
(\ref{eq:bbar-ttbar-partial-widths})) {\it i.e.}  $\epsilon_k/\mu$ not $\epsilon_i/\mu$. The $b\bar b$ mode, on the other hand, is
governed by $\epsilon_i/\mu$ (same flavor as the sneutrino) if $\epsilon_i\gg v_i$, or
$v_i/v$, if $\epsilon_i\ll v_i$.

Therefore, when $\epsilon_i\gg v_i$ (region where $bb$ modes dominate $WW$), neutrino
constraints imply all three $\epsilon$s are of the same order (see
Fig. \ref{fig:epsilons-vevs}). Therefore the different lepton flavor
final states become subdominant to the $bb$ modes due to charged lepton Yukawa
suppression. In contrast, when $\epsilon_i\ll v_i$ ($WW$ modes dominate $bb$ modes) the different
lepton flavor final states can have larger partial widths than the
$b\bar b$ mode provided $[(\epsilon_k/\mu)/(v_i/v)]^2$ exceeds the
corresponding Yukawa suppression $(h_i^E/h_b)^2$, which turns out to
be possible only for second and third generation sneutrinos as shown
in Fig. \ref{fig:epsilons-vevs}, where it can be seen that
$[(\epsilon_k/\mu)/(v_i/v)]^2$ can be at most $10^4$ (note
that $(h_\tau/h_b)^2\sim 10^{-1}$, $(h_\mu/h_b)^2\sim 10^{-3}$ while
$(h_e/h_b)^2\sim 10^{-8}$). As in the case of heavy SM and invisible
modes, these different lepton flavor final states can dominant over
the $b\bar b$ final state only in a single sneutrino generation.

\begin{figure}
  \centering
  \includegraphics[width=7.5cm,height=6.5cm]{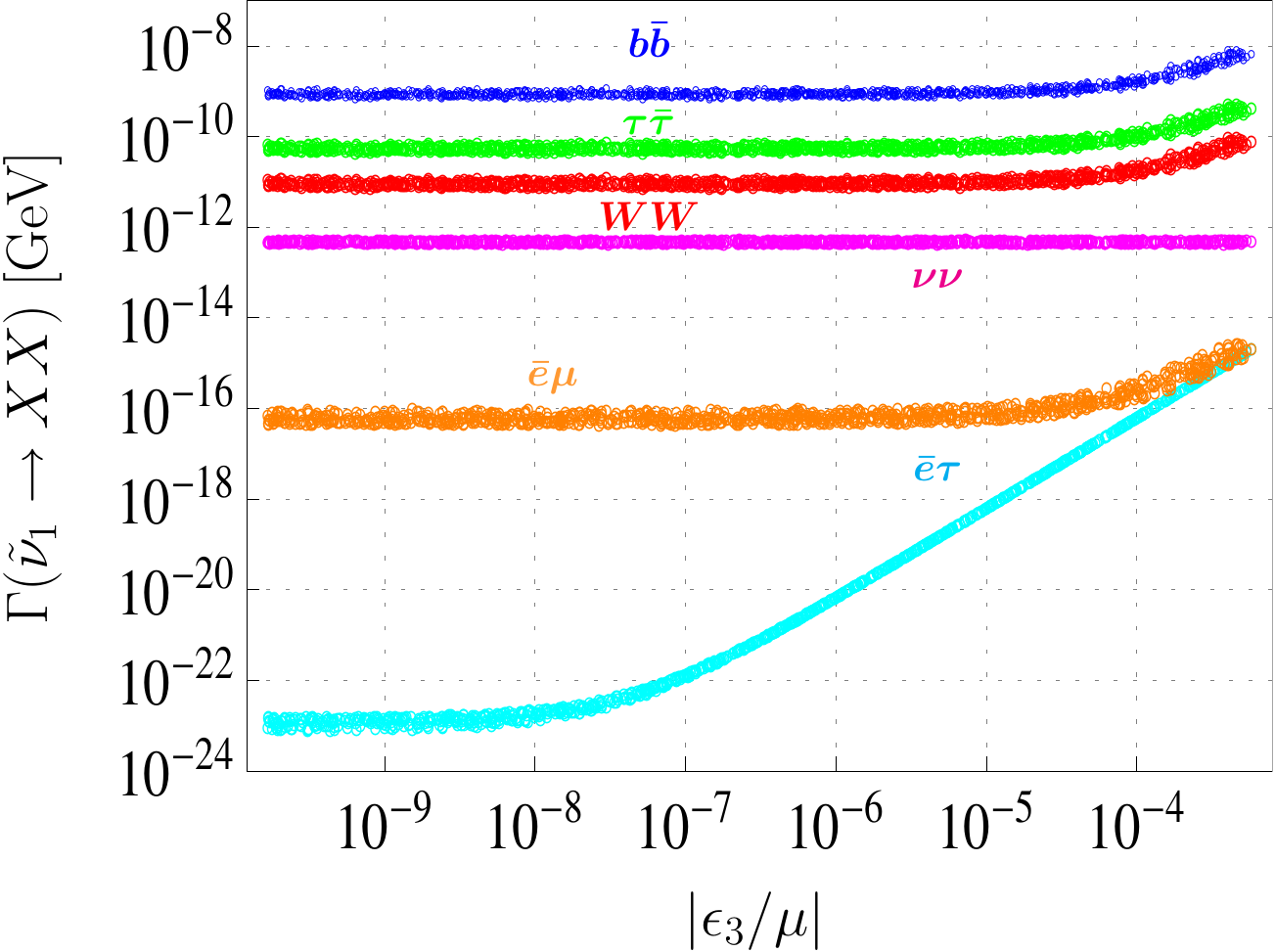}
  \hfill
  \includegraphics[width=7.5cm,height=6.7cm]{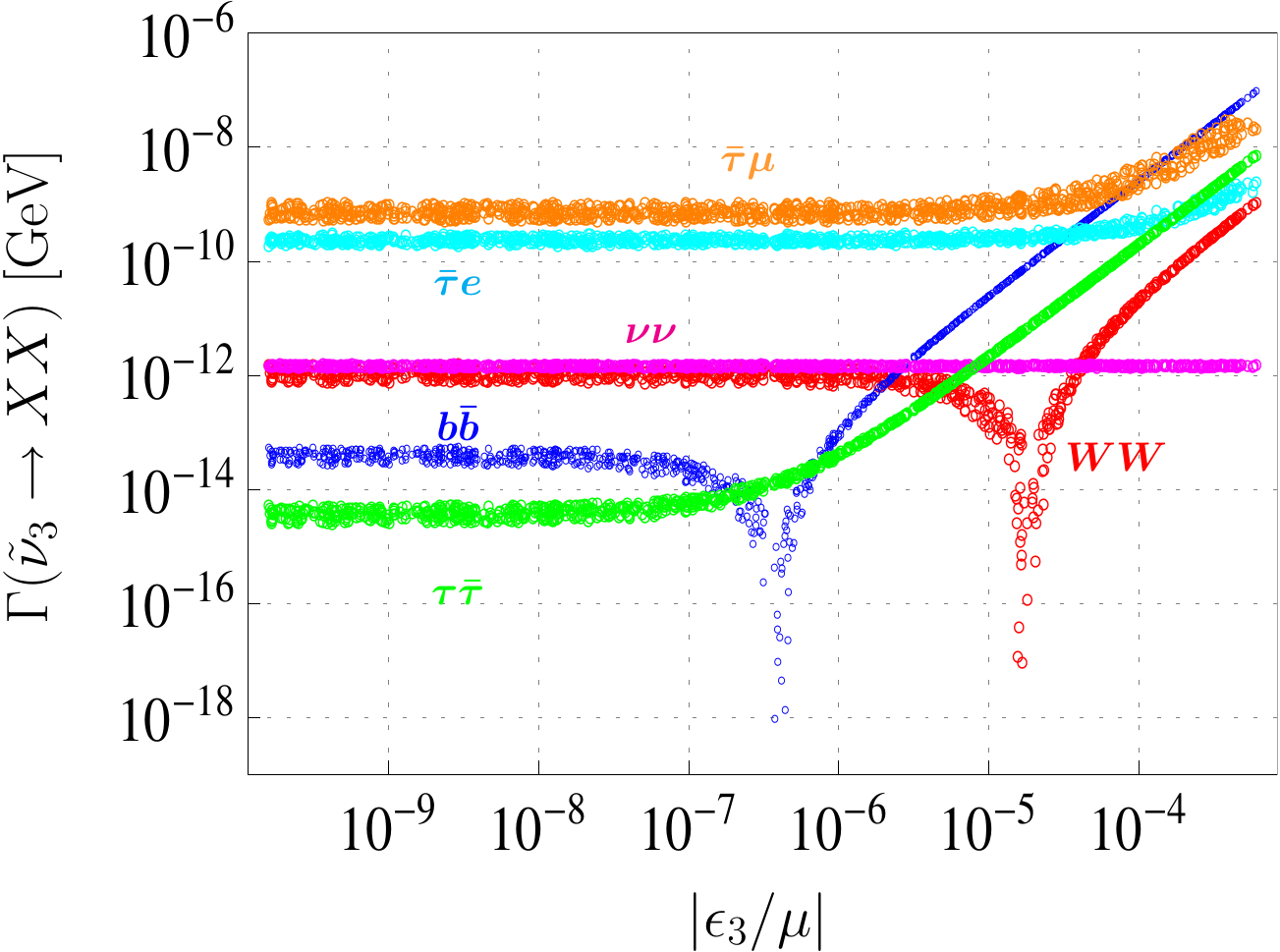}
  \caption{\it Partial decay widths for $\tilde\nu_{1}$ (left) and
    $\tilde\nu_{3}$ (right) to $b\bar b$, $WW$, $\tau\bar \tau$,
    $\sum_{j,k}\nu_j\nu_k$, $e\tau$, $e\mu$ and $\tau\mu$ final states
    as a function of $|\epsilon_3/\mu|$. BRpV parameters have been
    fixed by using neutrino oscillation data (see the text for more
    details). The remaining heavy SM modes, $hh$ and $ZZ$, have the
    same behavior and are of the same order than the $WW$ mode (are
    not shown to avoid crowding the plots). Decays of $\tilde \nu_2$
    are not shown since they are very similar to those of $\tilde
    \nu_1$.}
  \label{fig:width-snu1-snu3}
\end{figure}

We continue by considering the specific SUSY point discussed in the
last section and examining first and third generation sneutrino decays into $b \bar b$, $W^+
W^-$, $\sum_{j,k}\nu_j\nu_k$ and different lepton flavor modes
assuming that BRpV is solely responsible for neutrino masses. We have
also calculated the partial decay widths for the $h^0h^0$ and $Z^0Z^0$
final states, finding that they have a similar dependence on
$\epsilon_3/\mu$ and are of the same order as the $W^+W^-$ mode (the
$t \bar t$ channel is suppressed due to the off-shell top for this
sneutrino mass). The results are displayed in
figure~\ref{fig:width-snu1-snu3} versus $\epsilon_3/\mu$, where it can be
seen that $\tilde\nu_1$ exhibits the usual bahavior quoted in the
literature, that is to say, a dominant $b\bar b$ mode followed by a
slightly Yukawa suppressed $\tau\bar \tau$ mode and strongly
suppressed $\nu\nu$, $WW$, $\bar e\mu$ and $\bar e\tau$ final
states. The huge supression of the different lepton flavor mode, $\bar
e\tau$, in the small $\epsilon_3$ region is due to the combined effect
of the electron Yukawa suppression and the smallness of
$v_3/\epsilon_1$.  The $\bar e\mu$ final state being controlled by
$\epsilon_2/\mu$ involves only the electron Yukawa suppression, and so
it has a larger partial decay width than the $\bar e \tau$ mode.

The situation for
$\tilde\nu_3$, however, is quite different. In the
region of large $\epsilon_3$ it features a $b\bar b$ dominant mode and
apart from the $\bar{\tau} e$ mode it behaves like $\tilde\nu_1$. In
the region of small $\epsilon_3$, in contrast, the
``non-conventional'' modes $\bar \tau \mu$ ($\bar \tau e$), $\nu\nu$,
$W^+W^-$, $h^0 h^0$ and $ZZ$ have the largest partial widths by far,
greatly surpassing the $b\bar b$ final state. In particular, the different lepton
flavor channels exceed this mode by more than two-orders of magnitude,
thus drastically changing the sneutrino LSP phenomenology.

Some words are in order as regards the size of the $\tau\bar \tau$
partial width in $\tilde\nu_3$ decays. This decay mode is a
unequivocal signal of BRpV models, but its size is strongly dependent
on the nature of the bilinear parameters.  In R-parity violating
models satisfying the conditions $m_{H_d}^2=m_{L_i}^2$ and $B=B_i$ at
certain high energy scale, the bilinears can be rotated away. They
are, however, subsequently induced at the TeV scale by RGE running
effects \cite{deCarlos:1996yh,de Carlos:1996du,Nardi:1996iy} therefore
inducing the $\tau\bar \tau$ mode, but with a quite suppressed partial
width (typically several orders of magnitude smaller than that of the
$b\bar b$ channel). In a MSSM based analysis, as the one performed
here, where all the BRpV parameters are treated as free parameters
subject only to phenomenological constraints this is not necessarily
the case, as demonstrated in the right-hand side plot in
Fig. \ref{fig:width-snu1-snu3}.


Figure~\ref{fig:width-snu1-snu3} reinforces the arguments given in the
previous paragraphs. Reiterating: even if bilinear R-parity violation
accounts for neutrino data, the $b\bar b$ channel is not necessarily
the dominant decay mode. We have checked that for third and second
generation sneutrino LSPs the different lepton flavor final states
dominate ($\bar \tau e$, $\bar{\tau} \mu$ for $\tilde\nu_3$ and $\bar
\mu e$, $\bar \mu \tau$ for $\tilde \nu_2$) while first generation
sneutrinos decays are driven by invisible and heavy SM modes ($hh$ and
$WW$). As already pointed out and exemplified for $\tilde\nu_3$ in
Fig. \ref{fig:width-snu1-snu3}, these decays can dominate only in a
single sneutrino generation. So, observing at least two different
sneutrino flavors decaying to the aforementioned modes\footnote{In
  principle for sneutrinos originating from chargino decays the
  corresponding sneutrino flavor could be tagged by the associated lepton flavor \cite{Aristizabal
    Sierra:2004cy}.} (something that turns out to be viable if the
sneutrino mass splittings do not allow the heavier sneutrinos to decay
into the LSP) will prove that BRpV does not account for neutrino data,
because this would require $v_i\gg \epsilon_i$ for those generations,
which is not possible if neutrino physics is determined solely by
BRpV, see Fig.  \ref{fig:vevs-noatmospheric-nosolar}.

\begin{figure}
  \centering
  \includegraphics[width=7.5cm,height=6.5cm]{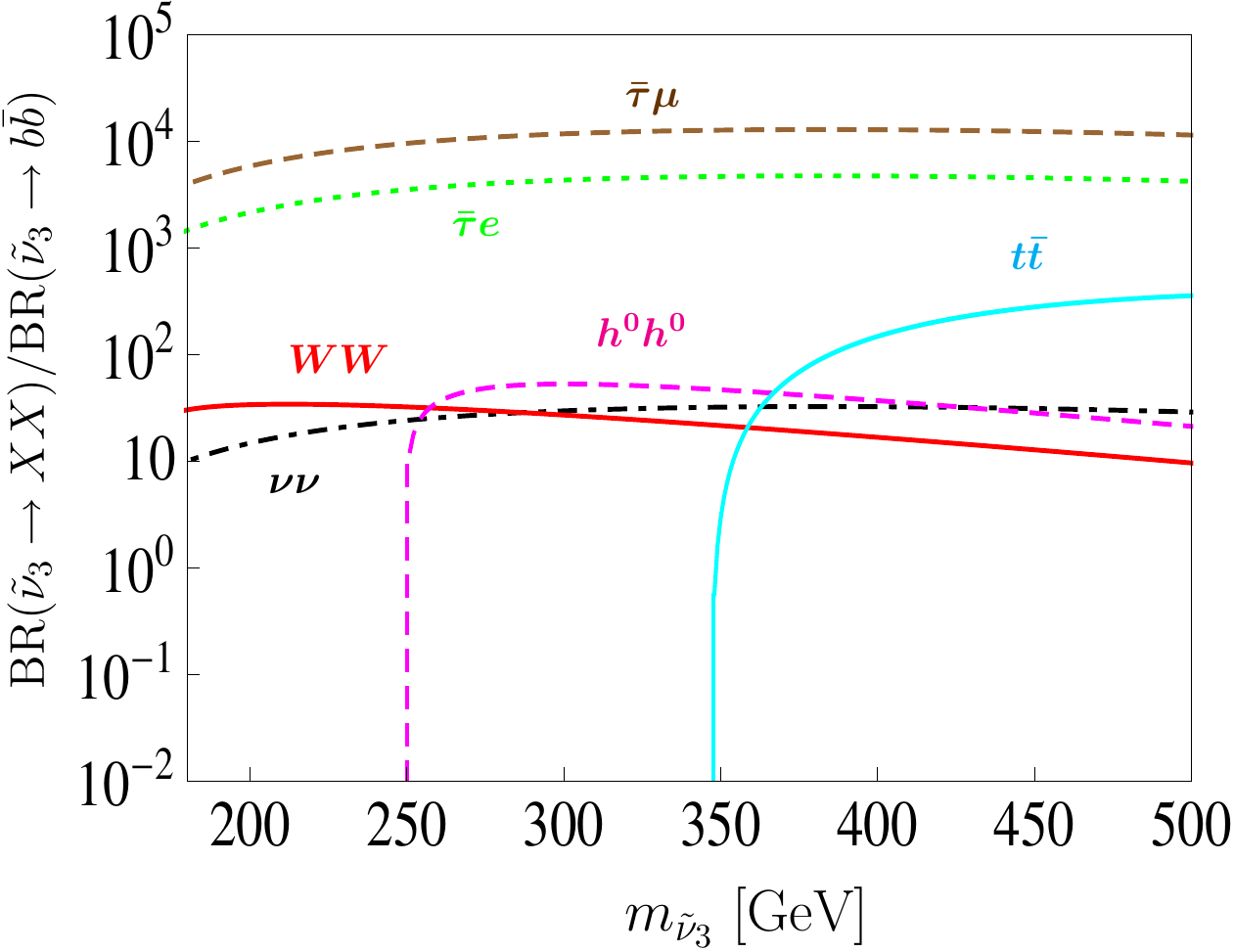}
  \includegraphics[width=7.5cm,height=6.5cm]{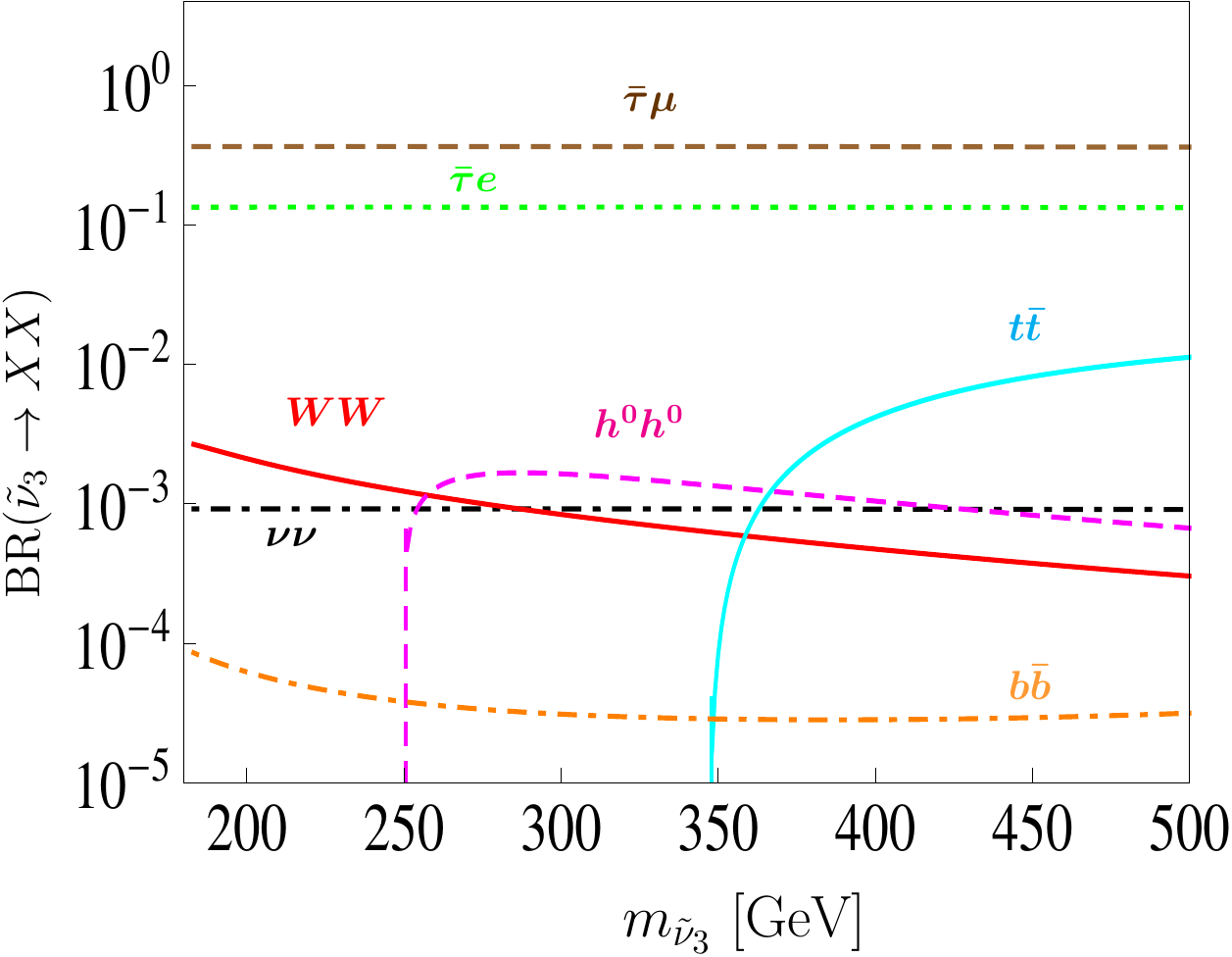}
  \caption{\it Left plot: decay branching fractions for different
    lepton flavor ($\bar \tau e$ and $\bar \tau \mu$), invisible
    ($\sum_{j,k}\nu_j\nu_k$), $WW$, $hh$ and $t\bar t$ final states
    normalized to $\mbox{BR}(\tilde\nu_3\to b\bar b)$ as a function of
    $m_{\tilde\nu_3}$. Right plot: decay branching ratios for dominant
    (excluding $ZZ$) and $b\bar b$ modes as a function of
    $m_{\tilde\nu_3}$.  The branching ratios have been calculated in
    the small $\epsilon_3$ region ($|\epsilon_3/\mu|=10^{-9}$) and for
    bilinear R-parity breaking parameters fixed by neutrino parameters
    according to their best fit point values
    \cite{Tortola:2012te,GonzalezGarcia:2012sz,Fogli:2012ua}. Both
    plots show the main result of this paper: sneutrino LSP decays are
    not necessarily dominated by the $b\bar b$ mode.}
  \label{fig:brxxoverbb-mnu}
\end{figure}

Focusing further on those final states of $\tilde \nu_3$ decays (excluding the $ZZ$ mode
since it behaves like the $WW$ mode and is smaller by an order of 1/2), we
study $\tilde \nu_3$ decays as a function of $m_{\tilde \nu_3}$ in
Figure~\ref{fig:brxxoverbb-mnu} allowing a $\tilde \nu_3$ heavy enough
so that the $t \bar t$ channel is also open. All the R-parity
conserving parameters have been fixed according to the supersymmetric
point used in the previous section. The R-parity breaking parameters
have been fixed by adjusting neutrino observables to their best fit
point values \cite{Tortola:2012te,GonzalezGarcia:2012sz,Fogli:2012ua}
and $\epsilon_3=\mu\times10^{-9}$. We have checked that the values are
insensitive to changes in $\tan\beta$ in the range
[2,30]. Off-shell calculations of the Higgs and top states have been
neglected since the different lepton flavor final states
dominant over the full sneutrino mass range. Both plots support our
claim: even if BRpV is the sole generator of neutrino masses a
sneutrino LSP does not necessarily decay predominantly into $b\bar b$
final states.

One can extrapolate from Figure 6 the behavior of the R-parity
violating sneutrino decays for the first two generations, in the
analogous region $\langle \tilde \nu_{1,2}\rangle \gg
\epsilon_{1,2}$. In these cases, only the leptonic different flavor
states shift and they do so by a factor of approximately
$(\boldsymbol{h^E}_\mu/\boldsymbol{h^E}_\tau)^2$ for $\tilde \nu_2 \to
\mu \bar \tau$ and $(\boldsymbol{h^E}_e/\boldsymbol{h^E}_\tau)^2$ for
$\tilde \nu_1 \to e \mu$. Therefore, $\tilde \nu_2$ decays will still
be dominated by different flavor leptonic final states, while $\tilde
\nu_1$ will decay predominately into heavy SM states and neutrinos.

\begin{figure}
  \centering
  \includegraphics[width=9cm,height=6.7cm]{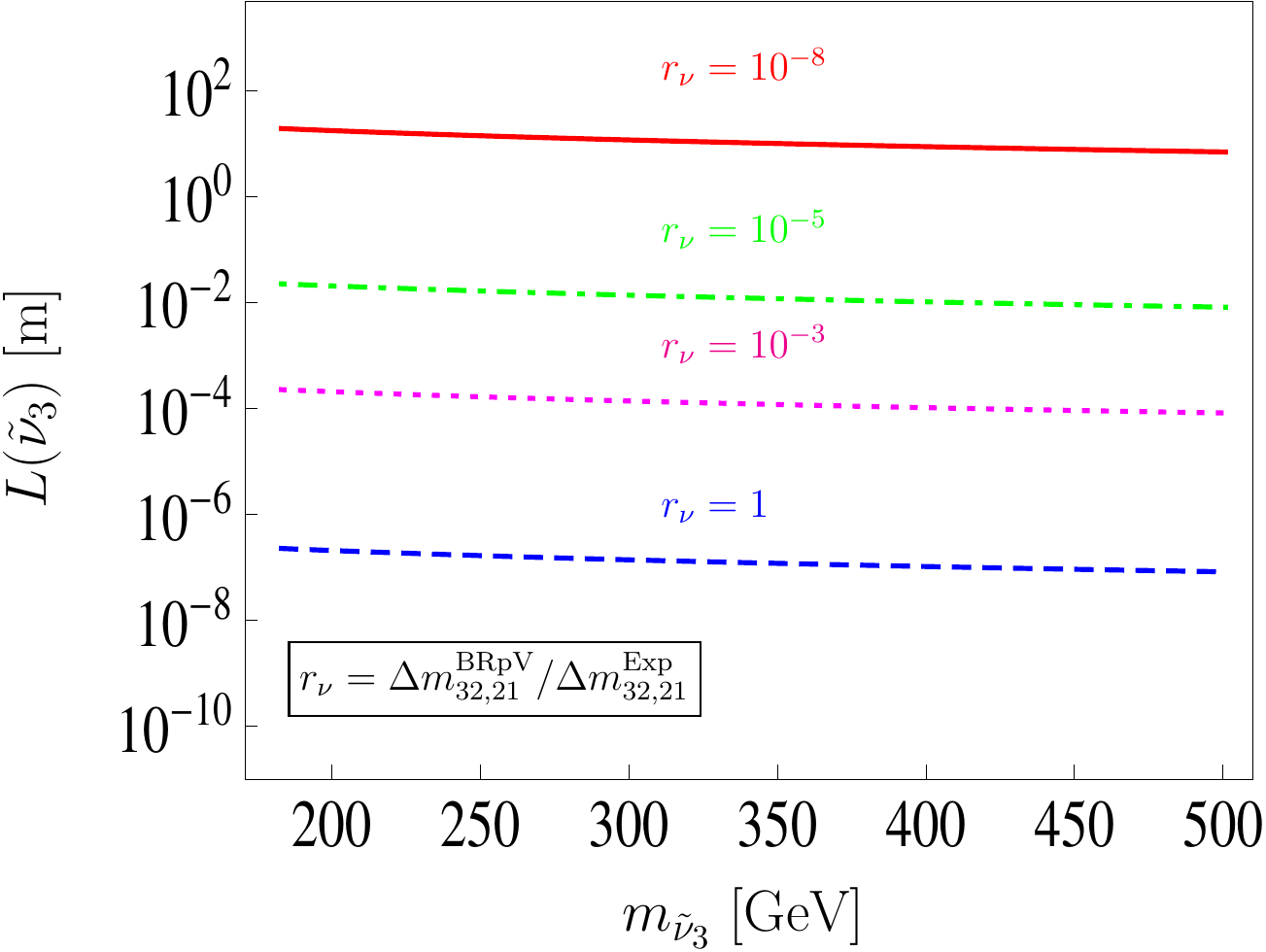}
  \caption{\it Sneutrino decay length for different choices of the
    bilinear R-parity breaking parameters.  The BRpV contribution to
    the atmospheric and solar mass scales is specified by the
    parameter $r_\nu\equiv \Delta m_{32,21}^\text{BRpV}/\Delta
    m^\text{Exp}_{32,21}$, where $\Delta m^\text{Exp}_{32,21}$ have
    been taken according to their best fit point values
    \cite{Tortola:2012te,GonzalezGarcia:2012sz,Fogli:2012ua}.}
  \label{fig:decay-length}
\end{figure}

Relaxing the assumption that bilinear R-parity violation is the sole
generator of neutrino masses (as in the second case
discussed in section~\ref{sec:constraints-neu-physics}) will not
drastically change the results in this section. This is because the relative
importance of the ratios
$\mbox{BR}^{\tilde\nu_i}_{XX}/\mbox{BR}^{\tilde\nu_i}_{b\bar b}$ is
determined by $v_i\gg\epsilon_i$. However, a quantity
that becomes relevant in this case is the sneutrino decay length $L(\tilde\nu_i)$,
since the size of the bilinear R-parity breaking
parameters can be small enough to cause the sneutrino to decay outside of the
detector. Neglecting the Lorentz boost factor, we have found that as
long as the R-parity violating parameters fit the neutrino data, in general
$L(\tilde\nu_i)$ is well below 1 mm (inline
with~\cite{Hirsch:2003fe}). We have also checked that if $r_\nu=\Delta
m_{32,21}^{\text{BRpV}}/ \Delta m^{\text{Exp}}_{32,21}= 10^{-5}$ the
decay length is generically below $\sim 10$ cm. Note that values of
$r_\nu$ in the range $[10^{-3},1]$ might be in conflict with neutrino
data when the contributions from the mechanism responsible for
neutrino masses are taken into account (for example a standard
seesaw). Once the contributions to the atmospheric and solar mass
scales fall below $r_\nu=10^{-5}$ the decay length can eventually
exceed 1 m almost independently of the sneutrino mass (as expected due
to the large partial decay widths the different lepton flavor modes
have). Figure~\ref{fig:decay-length} shows the decay length for
$r_\nu=1,10^{-3},10^{-5},10^{-8}$.

Thus far, this section has focused on sneutrino decays. At this point,
we briefly comment on production. If the sneutrino is indeed the LSP,
all produced supersymmetric particles produced will cascade decay
down to it. The dominate production depends on the relative masses,
with strong production~\cite{Kramer:2012bx} dominating unless the gluinos and squarks are too
heavy, in which case relatively light sneutrinos can still be pair produced
through electroweak processes. Regardless, the parameter region explored
in this paper yields compelling signals since the dominate different flavor leptonic final states
would have little background if the intermediate sneutrino can be reconstructed.
Meanwhile, the $b\bar b$ mode corresponding to the parameter space typically
considered in the literature would suffer from larger QCD background and would therefore
be more difficult to detect. Furthermore, if the BRpV
contributions to neutrino masses are subdominant to some other mechanism, sneutrino decays
could involve large displaced vertices (see Fig. \ref{fig:decay-length})
thus rendering its identification even more plausible.

\section{Conclusions}
\label{sec:conclusions}
BRpV is a good effective theory for models of spontaneous R-parity
violation, in which proton decay as well as the number of new
parameters is under control compared to explicit R-parity violation.
In the presence of BRpV, sneutrino LSP decays can in some cases
resemble a heavy Higgs. However, as we have discussed, sneutrinos
(third and second generation) can dominantly decay to different-flavor
charged leptons final states or heavy SM modes (first generation),
thus completely altering the ``conventional'' sneutrino LSP
phenomenology driven by the $b\bar b$ mode.

We have studied these decays into $W^+W^-$, $ZZ$, $h^0h^0$, $t\bar t$,
$\nu\nu$ and $l_i^+l_k^-$ ($i\neq k$) and found that long as
$\epsilon_i\ll v_i$, they have larger branching ratios than the $b\bar
b$ mode. As discussed in section~\ref{sec:constraints-neu-physics},
for models where the BRpV parameters are fixed by neutrino data, this is possible only for a
single sneutrino flavor, due
to the constraints from the solar sector. Therefore, if 2 or more sneutrino generations
are degenerate enough to decay via BRpV into these states one could
rule out BRpV as the sole source of neutrino masses.

For models where the bilinear R-parity breaking parameters do not
contribute significantly to neutrino masses, the constraint
$\epsilon_i \ll v_i$ for all flavors is viable. Accordingly, in these
BRpV models large sneutrino LSP branching fractions into either
$l_i^+l_k^-$ or $W^+W^-$, $ZZ$, $h^0h^0$ and $t\bar t$ are
possible. In this case, however, special attention has to be paid to
the sneutrino decay length $L(\tilde\nu_i)$. We calculated
$L(\tilde\nu_i)$, finding that as long as the BRpV contributions to
neutrino masses are not below $\sim 10^{-3}\%$ the decay length is
generically below $\sim 10$ cm.

Regardless of these considerations, sneutrino decays into
different-flavor charged leptons or heavy SM final states
($l_i^+l_k^-$ or $W^+W^-$, $ZZ$, $h^0h^0$ and $t\bar t$) are an
interesting phenomenological possibility that has not received much
attention before. Because BRpV only affects the decays of the LSP,
assuming a sneutrino LSP means that SUSY events could cascade decay to
either different flavor charged leptons or two pairs of heavy SM states, which
constitutes a unique and novel SUSY signature.

\section{Acknowledgments}
We want to thank Martin Hirsch for his comments in the early stage of
this project and Werner Porod for SPheno support. SS would also like
to thank Pavel Fileviez Perez for comments. DAS is supported by a
Belgian FNRS postdoctoral fellowship. DR has been partially supported
by Sostenibilidad-UdeA and COLCIENCIAS through the grant number
111556934918.


\end{document}